\documentclass[12pt,a4paper]{article}
\usepackage[OT2,T1]{fontenc}
\usepackage{amssymb}
\usepackage{amsmath}
\usepackage{stmaryrd}
\usepackage{mathrsfs}
\usepackage{mathtools}
\usepackage{bm}
\usepackage{graphicx}
\usepackage{here}
\usepackage{subcaption}
\usepackage[margin=1in]{geometry}
\usepackage{cite}
\usepackage{sectsty}
\allsectionsfont{\sffamily}
\usepackage[colorlinks,allcolors=blue]{hyperref}
\DeclareMathAlphabet{\mathscrbf}{OMS}{mdugm}{b}{n}
\numberwithin{equation}{section} % numbers equations as (sectionNumber.equationNumber)

\let\OLDthebibliography\thebibliography
\renewcommand\thebibliography[1]{
  \OLDthebibliography{#1}
  \setlength{\parskip}{0pt}
  \setlength{\itemsep}{0.5pt plus 0.3ex}
}

\newcommand{\be}{\begin{equation}}
\newcommand{\ee}{\end{equation}}
\newcommand{\ben}{\begin{enumerate}}
\newcommand{\een}{\end{enumerate}}
\newcommand{\bi}{\begin{itemize}}
\newcommand{\ei}{\end{itemize}}
\newcommand{\bmm}{\begin{pmatrix}}
\newcommand{\emm}{\end{pmatrix}}

\newcommand{\Ad}{\text{Ad}}
\newcommand{\bra}{\langle}

\newcommand{\demi}{\frac{1}{2}}
\newcommand{\der}{\partial}

\newcommand{\ds}{\displaystyle}
\newcommand{\eg}{e.g.\ }

\newcommand{\ie}{i.e.\ }
\newcommand{\ket}{\rangle}
\newcommand{\nn}{\nonumber}

\renewcommand{\refeq}[1]{\stackrel{\text{(\ref{#1})}}{=}}

\newcommand{\balpha}{{\boldsymbol\alpha}}

\newcommand{\bbeta}{{\boldsymbol\beta}}
\newcommand{\bk}{\textbf k}
\newcommand{\bbq}{\textbf q}

\newcommand{\bv}{\textbf v}
\newcommand{\bw}{\textbf w}
\newcommand{\bx}{\textbf x}
\newcommand{\by}{\textbf y}

\newcommand{\cA}{{\cal A}}

\newcommand{\cF}{{\cal F}}

\newcommand{\cO}{{\cal O}}

\newcommand{\cS}{{\cal S}}

\newcommand{\cU}{\,{\cal U}}
\newcommand{\cW}{{\cal W}}

\newcommand{\mg}{\mathfrak{g}}
\newcommand{\mh}{\mathfrak{h}}
\newcommand{\cs}{\mathfrak{s}}
\newcommand{\cu}{\mathfrak{u}}

\newcommand{\sH}{\mathscr{H}}

\newcommand{\sfW}{\mathsf{W}}
\newcommand{\sfw}{\mathsf{w}}

\newcommand{\CC}{\mathbb{C}}

\newcommand{\II}{\mathbb{I}}

\newcommand{\RR}{\mathbb{R}}

\begin{document}

\hrule
\begin{center}
\Large{\bfseries{\textsf{Probing Wigner Rotations for Any Group}}}
\end{center}
\hrule
~\\

\begin{center}
\large{\textsf{Blagoje Oblak$^{*}$}}
\end{center}
~\\

\begin{center}
\begin{minipage}{.9\textwidth}\small\it
\begin{center}
Institut f\"ur Theoretische Physik\\
ETH Z\"urich\\
CH-8093 Z\"urich, Switzerland
\end{center}
\end{minipage}
\end{center}

\vspace{2cm}

\begin{center}
\begin{minipage}{.9\textwidth}
\begin{center}{\bfseries{\textsf{Abstract}}}\end{center}
Wigner rotations are transformations that affect spinning particles and cause the observable phenomenon of Thomas precession. Here we study these rotations for arbitrary symmetry groups with a semi-direct product structure. In particular we establish a general link between Wigner rotations and Thomas precession by relating the latter to the holonomies of a certain Berry connection on a momentum orbit. Along the way we derive a formula for infinitesimal, Lie-algebraic transformations of one-particle states.
\end{minipage}
\end{center}

\vfill
\noindent
\mbox{}
\raisebox{-3\baselineskip}{%
\parbox{\textwidth}{\mbox{}\hrulefill\\[-4pt]}}
{\scriptsize$^*$ E-mail: boblak@phys.ethz.ch.}

\thispagestyle{empty}
\newpage

\textsf{\tableofcontents}
\thispagestyle{empty}

\newpage
\setcounter{page}{1}
\pagenumbering{arabic}

%%%%%%%%%%%%%%%%%%%%%%%%%%%%%%%%%%%%%%%%%%%%%%%%%%%%%%%%%%%
\section{Introduction}
%%%%%%%%%%%%%%%%%%%%%%%%%%%%%%%%%%%%%%%%%%%%%%%%%%%%%%%%%%%

Our understanding of particle physics relies to a great extent on the properties of the Poincar\'e group --- the isometry group of Minkowski space-time, comprising Lorentz transformations and space-time translations. In that context, irreducible unitary representations of Poincar\'e are interpreted as relativistic particles. They can be classified according to their mass and spin, as shown long ago by Wigner \cite{Wigner:1939cj}. However, recent insights in high-energy physics and general relativity hint that the Poincar\'e group is not that fundamental after all. Indeed, about fifty years ago, Bondi, van der Burg, Metzner and Sachs found that Poincar\'e symmetry is obsolete in space-times with a gravitational field, and gets enhanced to an infinite-dimensional asymptotic symmetry now known as the BMS group \cite{Bondi:1962px}. Shortly thereafter, it was suggested that one could define a corresponding generalized notion of particles, and their classification was performed by McCarthy \cite{McCarthy01}. The last few years have witnessed a resurgence of interest in BMS symmetry, as it was argued that it can be extended even further \cite{Barnich:2009se}, that it reproduces various soft theorems in quantum field theory \cite{Strominger:2013lka}, and that it might account for black hole entropy \cite{Hawking:2016msc}.\\

The structure of the BMS group is similar to that of Poincar\'e: it is a semi-direct product containing a non-Abelian group of Lorentz transformations or `superrotations' acting on a vector group of `supertranslations'. This constrains all unitary representations of BMS \cite{Barnich:2014kra,Barnich:2015uva}, which in turn constrains all quantum-mechanical systems with BMS symmetry. It is therefore of interest to study how various group-theoretic observables change when enhancing Poincar\'e to BMS. The present work originates from an attempt to describe one such observable, namely {\it Thomas precession} \cite{Thomas:1926dy}, in the BMS context. While investigating this topic, it turned out that many of the required tools were not readily available in the literature. Indeed, it is widely known that Thomas precession follows from the presence of {\it Wigner rotations} in the transformation law of one-particle states, but most references on this subject in the physics literature only treat it for the very special case of Poincar\'e symmetry; see \eg \cite[sec.\ 2.5]{Weinberg:1995mt}, \cite[sec.\ 11.8]{Jackson:1998nia} or the papers \cite{Ferraro:1999eu,Dragan:2012ir,Ungar1988}. On the other hand, the mathematics literature mostly seems to focus on abstract structures rather than the concrete computations needed by physicists; see \eg \cite[sec.\ V.5]{varadarajan1968geometry} or \cite{Ungar1991}. (Wigner rotations are also responsible for entanglement between spin and momentum \cite{Peres:2002ip}, but we will not investigate this here.)\\

The purpose of this paper is to fill this gap and describe, in full generality, various aspects of Wigner rotations in unitary representations of semi-direct products. In short, the question we wish to address is the following: given any semi-direct product group, are the corresponding particles subject to Thomas precession? And if yes, what is the precession rate? We shall formulate the answer in terms of a Berry connection (\ref{ss25b}) related to the Maurer-Cartan form of the symmetry group. Along the way we will derive eq.\ (\ref{ss47}) for unitary Lie algebra representations of semi-direct sums; surprisingly, we were unable to find this formula in the literature, so we hope it can be useful in a broader context than that of this work. Throughout the paper, we illustrate our results with the Poincar\'e group (which does display Thomas precession) and the Bargmann group (which does not). The BMS group will be treated in a separate publication \cite{OblakThomas}.\\

The plan is as follows. We start in section \ref{sec2} by recalling the Wigner-Mackey construction of irreducible unitary representations of semi-direct products, with particular emphasis on Wigner rotations and their cohomological properties; we also illustrate this method with massive representations of the Poincar\'e group (relativistic particles) and the Bargmann group (non-relativistic particles). Section \ref{sec4} is devoted to the Lie algebra representations obtained by differentiating the Wigner-Mackey formula; these are then used in section \ref{sec5} to describe Thomas precession as a Berry phase in a Hilbert space with a continuous energy spectrum. Finally, in section \ref{sec6} we apply these results to the Poincar\'e and Bargmann groups and briefly discuss other potential applications.\\

A disclaimer may be called for before we start. The present work was originally meant as a technical appendix to \cite{OblakThomas}, but it eventually turned out that the resulting structures were interesting enough by themselves (at least in the author's opinion) to deserve a paper of their own. We hope that the reader will not be put off by the abstractness of our presentation. This being said, note that our approach will not be mathematically rigorous, so all necessary smoothness or regularity conditions are tacitly assumed to hold.

%%%%%%%%%%%%%%%%%%%%%%%%%%%%%%%%%%%%%%%%%%%%%%%%%%%%%%%%%%%
\section{Wigner rotations}
\label{sec2}
%%%%%%%%%%%%%%%%%%%%%%%%%%%%%%%%%%%%%%%%%%%%%%%%%%%%%%%%%%%

Here we review the description of one-particle states as induced representations of semi-direct products, based on the notion of orbits and little groups. For spinning particles, this involves Wigner rotations that we describe in detail. We refer \eg to \cite[chap.\ 16-17]{barut1986theory} or \cite[chap.\ 4]{Oblak:2016eij} for an introduction to these matters with a milder learning curve.

%%%%%%%%%%%%%%%%%%%%%%%%%%%%%%%%%%%%%
\subsection{Orbits, little groups and standard boosts}
\label{sec21}
%%%%%%%%%%%%%%%%%%%%%%%%%%%%%%%%%%%%%

Consider a Lie group $G$, generally non-Abelian, whose elements we write as $f$, $g$, etc. Let also $A$ be a vector space with elements $\alpha$, $\beta$, etc.; one can think of $A$ as an Abelian group with respect to vector addition. Finally, let $\sigma$ be a representation of $G$ in $A$, so that for each $f\in G$ we have a linear operator $\sigma_f$ acting on $A$. Then the {\it semi-direct product} of $G$ and $A$ is the group $G\ltimes A$ whose elements are pairs $(f,\alpha)$, with a group operation
\be
(f,\alpha)\cdot(g,\beta)
=
\big(f g,\alpha+\sigma_f\beta\big).
\label{s13}
\ee
Many interesting groups in physics are semi-direct products. Examples include the Poincar\'e groups, the Galilei groups (and their central extensions, the Bargmann groups), as well as the BMS groups. In all these cases, the space $A$ is interpreted as a group of translations while $G$ consists of rotations or boosts that act on $A$.\\

Given a semi-direct product $G\ltimes A$, let $A^*$ be the dual vector space of $A$. In keeping with the interpretation of $A$ as a group of translations, we shall think of $A^*$ as `momentum space' and denote its elements as $p$, $q$, etc. Each momentum $p$ is a linear form on $A$, $\langle p,\cdot\rangle:A\rightarrow\RR:\alpha\mapsto\langle p,\alpha\rangle$. The action $\sigma$ of $G$ on $A$ gives rise to its dual action on momenta: for each $f\in G$ and any $p\in A^*$, we define $f\cdot p\in A^*$ by
\be
\langle f\cdot p,\alpha\rangle
\equiv
\langle p,\sigma_{f^{-1}}\alpha\rangle
\qquad\forall\,\alpha\in A.
\label{s15}
\ee
Choosing a momentum vector $p$, we define its {\it orbit} under $G$ as
\be
\cO_p\equiv\big\{f\cdot p\,\big|\,f\in G\big\}.
\label{s16}
\ee
It is the set of all momenta that can be reached by acting on $p$ with $G$. Not all elements of $G$ act non-trivially on $p$; those that leave it fixed span the {\it little group} of $p$,
\be
G_p
=
\big\{
f\in G
\,\big|\,
f\cdot p=p
\big\}.
\label{ss16}
\ee
For the Poincar\'e group, the dot action given by (\ref{s15}) is just the transformation law of energy-momentum vectors under the Lorentz group and each orbit (\ref{s16}) is a hyperboloid specified by an equation of the type $q_{\mu}q^{\mu}=-M^2$ \cite{Wigner:1939cj}. For the BMS$_3$ group, the action of $G$ on (super)momenta is that of (chiral) conformal transformations on CFT stress tensors and each orbit is a coadjoint orbit of the Virasoro group \cite{Barnich:2014kra,Barnich:2015uva}.\\

Let us now pick one particular momentum orbit $\cO_p$. By construction, the action of $G$ on the orbit is transitive, so $\cO_p$ is diffeomorphic to the quotient space $G/G_p$ and for any $q\in\cO_p$ we can find a group element $g_q$ such that
\be
g_q\cdot p=q.
\label{s17}
\ee
We shall assume that the $g_q$'s are chosen so as to depend smoothly on $q$ over the entire orbit\footnote{This assumption is tantamount to the triviality of the $G_p$-bundle $G\rightarrow\cO_p$. It is not satisfied in general, but it will hold for all massive Poincar\'e or Bargmann orbits considered below.}, and refer to them as a family of {\it standard boosts}. They can be seen as a map
\be
g:\cO_p\rightarrow G:q\mapsto g_q
\label{s17b}
\ee
which is in fact a section of the $G_p$-bundle $G\rightarrow\cO_p$ by virtue of eq.\ (\ref{s17}). Note that standard boosts are not uniquely defined: given some $g_q$'s satisfying (\ref{s17}), we can always multiply them from the right by any family of little group elements $h_q\in G_p$ without affecting the requirement (\ref{s17}). One can think of the mapping
\be
g_q\mapsto g_qh_q
\label{ss17}
\ee
as a gauge transformation on the orbit, with gauge group $G_p$.

%%%%%%%%%%%%%%%%%%%%%%%%%%%%%%%%%%%%%
\subsection{Wigner rotations and one-particle states}
\label{sec22}
%%%%%%%%%%%%%%%%%%%%%%%%%%%%%%%%%%%%%

The notions of orbits, little groups and standard boosts are at the core of the irreducible representations of semi-direct products built by Wigner and Mackey \cite{Wigner:1939cj,Mackey01}. The construction goes as follows: pick an orbit $\cO_p$ and let $\cS$ be an irreducible unitary representation of $G_p$. The choice of $\cO_p$ will eventually determine the allowed momenta of one-particle states (in Poincar\'e it fixes the value of the mass parameter), while $\cS$ determines their spin. We shall refer to the carrier space of $\cS$ as the `spin space' and denote it by $\mh$. Then any wavefunction $\Psi$ representing a one-particle state is a map
\be
\Psi:\cO_p\rightarrow\mh:q\mapsto\Psi(q).
\label{s18}
\ee
One can think of it as a wavefunction in the momentum picture of quantum mechanics. All wavefunctions are required to be square-integrable and their scalar products read
\be
\big<\Phi\big|\Psi\big>
=
\int_{\cO_p}d\mu(q)\big(\Phi(q)\big|\Psi(q)\big)
\label{ss19}
\ee
where $\mu$ is a measure on $\cO_p$ while $(\cdot|\cdot)$ denotes the scalar product in the spin space $\mh$. Thus the Hilbert space of one-particle states is $\sH=L^2(\cO_p)\otimes\mh$. The choice of $\mu$ is mostly irrelevant, but for simplicity we shall assume that it is invariant under $G$ so that $d\mu(f\cdot q)=d\mu(q)$ for any $f\in G$.\footnote{This assumption can be relaxed by allowing $\mu$ to be {\it quasi}-invariant under $G$ (see \eg \cite{Oblak:2015sea}), but this complication affects the resulting representations only mildly so we do not include it here.} The action $\cU$ of $G\ltimes A$ on any wavefunction (\ref{s18}) is
\be
\big(\cU[(f,\alpha)]\cdot\Psi\big)(q)
=
e^{i\bra q,\alpha\ket}\cS[g_q^{-1}fg_{f^{-1}\cdot q}]\cdot\Psi(f^{-1}\cdot q)
\label{s19}
\ee
where the $g_q$'s are standard boosts. One can show that $\cU$ is an irreducible, unitary representation of $G\ltimes A$, and also that {\it any} such representation takes the form (\ref{s19}) for some unique choice of $\cO_p$ and $\cS$ \cite{mackey1949imprimitivity}. In this sense, the one-particle states of any semi-direct product are always uniquely labelled by their `mass' $\cO_p$ and `spin' $\cS$. In what follows we refer to this statement as the {\it Wigner-Mackey theorem}.\\

In eq.\ (\ref{s19}), the exponential $e^{i\bra q,\alpha\ket}$ represents the usual action of translations on wavefunctions in momentum space, while the argument $f^{-1}\cdot q$ of $\Psi$ on the right-hand side accounts for the transformation law of scalar wavefunctions under boosts and rotations. But the term involving $\cS$ is more intriguing: it is a {\it Wigner rotation}\footnote{The Wigner rotations introduced here should not be confused with the `Wigner rotation matrices' or `Wigner $D$-matrices' appearing in representations of $\text{SU}(2)$, as these notions are completely unrelated.}
\be
\sfW_q[f]
\equiv
\cS\big[\,g_q^{-1}\,f\,g_{f^{-1}\cdot q}\,\big]
\label{s20}
\ee
that contains a carefully crafted combination of group elements, designed in just the right way to belong to the little group (\ref{ss16}). Indeed, using the defining property (\ref{s17}) of standard boosts, one finds that the combination leaves $p$ invariant:
\be
g_q^{-1}fg_{f^{-1}\cdot q}\cdot p
=
g_q^{-1}f\cdot f^{-1}\cdot q
=
g_q^{-1}\cdot q
=
p.
\nn
\ee
Intuitively, the Wigner rotation operator (\ref{s20}) represents the action of the transformation $f$ on the spin of a particle with momentum $f^{-1}\cdot q$. 

%%%%%%%%%%%%%%%%%%%%%%%%%%%%%%%%%%%%%
\subsection{Gauge invariance and cohomology}
\label{sec23}
%%%%%%%%%%%%%%%%%%%%%%%%%%%%%%%%%%%%%

Wigner rotations satisfy several important properties. First, they are not invariant under gauge transformations (\ref{ss17}), since mapping $g_q$ on $g_qh_q$ transforms the operator (\ref{s20}) as
\be
\sfW_q[f]
\mapsto
\cS[h_q]^{-1}\cdot\sfW_q[f]\cdot\cS[h_{f^{-1}\cdot q}].
\label{s21}
\ee
Thus, different choices of standard boosts give different Wigner rotations, which is to say that the representation (\ref{s13}) depends on the $g_q$'s. Nevertheless, two such representations with identical orbit and spin but different standard boosts are unitarily equivalent, as the change (\ref{s21}) leaves the representation (\ref{s19}) invariant provided one rewrites it in terms of $\cS[h_q]\Psi(q)$ rather than $\Psi(q)$. In this sense the choice of $g_q$'s is merely a gauge choice.\\

The gauge-dependence of Wigner rotations implies that they cannot be observed directly in any experiment. This naively suggests that they may be removed altogether, \ie that the transformation (\ref{s19}) could just as well be written without explicit reference to $\cS$. But this cannot be true, since it would lead to the absurd conclusion that any particle with spin $\cS$ is equivalent to a collection of $\text{dim}(\cS)$ scalar particles. So in fact, Wigner rotations {\it cannot} be bluntly removed, and the problem becomes to extract gauge-invariant observables out of gauge-dependent Wigner rotations. How can that be done? As it turns out, the answer will be provided by Thomas precession.\\

Wigner rotations are also amenable to cohomological considerations. Indeed, one can think of them as maps that send a point $q\in\cO_p$ on an operator $\sfW_q[f]\in\text{End}(\mh)$ given by eq.\ (\ref{s20}). If we denote the space of all such (smooth) maps by $C^{\infty}\big(\cO_p,\text{End}(\mh)\big)$, then Wigner rotations define an assignment
\be
\sfW:G\rightarrow C^{\infty}\big(\cO_p,\text{End}(\mh)\big):f\mapsto\sfW[f]
\label{s23b}
\ee
where $\sfW[f]$ is an $\text{End}(\mh)$-valued function on $\cO_p$ whose value at $q$ is $\sfW_q[f]$. Now note that, for $f,g\in G$, the definition (\ref{s20}) implies\footnote{Abstractly, (\ref{s23}) says that Wigner rotations provide a representation of the action groupoid $G\ltimes\cO_p$.}
\be
\sfW_q[fg]
=
\sfW_q[f]\cdot\sfW_{f^{-1}\cdot q}[g].
\label{s23}
\ee
In cohomological terms, this states that the map (\ref{s23b}) is a one-cocycle on $G$ \cite[sec.\ V.5]{varadarajan1968geometry}. This is most manifest when the spin space $\mh=\CC$ is one-dimensional; then any Wigner rotation is an exponential
\be
\sfW_q[f]
=
e^{i\cW_q[f]}
\label{t23}
\ee
with a real phase $\cW_q[f]$, and eq.\ (\ref{s23}) means that $\cW_q[fg]=\cW_q[f]+\cW_{f^{-1}\cdot q}[g]$. Here $\cW[f]$ is a function that maps $q\in\cO_p$ on the number $\cW_q[f]$. The group $G$ acts on any such function $\phi$ as $\big(f\cdot\phi\big)(q)=\phi(f^{-1}\cdot q)$, so one can write $\cW[fg]=\cW[f]+f\cdot\cW[g]$, which precisely says that the map $\cW:f\mapsto\cW[f]$ is a one-cocycle on $G$. Futhermore, if this cocycle is trivial so that $\cW_q[f]=\phi(f^{-1}\cdot q)-\phi(q)$ for some function $\phi$, then the representation (\ref{s19}) can be rewritten as
\be
\big(\cU[(f,\alpha)]\cdot e^{i\phi}\Psi\big)(q)
=
e^{i\bra q,\alpha\ket}\big(e^{i\phi}\Psi\big)(f^{-1}\cdot q),
\nn
\ee
which is to say that the exponential (\ref{t23}) can be absorbed by a redefinition of $\Psi$. This also applies when the spin space is not one-dimensional: in that case one would say that Wigner rotations are cohomologically trivial if they can be written as $\sfW_q[f]=\Omega(q)^{-1}\cdot\Omega(f^{-1}\cdot q)$ for some function $\Omega$ on $\cO_p$ valued in $\text{GL}(\mh)$. Such Wigner rotations can be removed from the representation (\ref{s19}) by expressing it in terms of $\Omega\cdot\Psi$ rather than $\Psi$.\\

Thus, representations of $G\ltimes A$ with cohomologically trivial Wigner rotations are scalar representations in disguise. But we stress that, by construction, the actual Wigner rotation (\ref{s20}) is {\it never} trivial (at least as long as $\cS$ is non-trivial). Indeed, the whole point of the Wigner-Mackey construction (\ref{s19}) is that different spins automatically specify inequivalent irreducible representations; if the Wigner rotations (\ref{s20}) were trivial for some irreducible choice of $\cS$, then the corresponding representations would be reducible, which would contradict the Wigner-Mackey theorem. In a way, the theorem is a recipe for building the only possible non-trivial cocycles of this type: all of them are given by (\ref{s20}) for some choice of spin. Note also that in cohomological language, eq.\ (\ref{s21}) states that Wigner rotations change by a coboundary when changing standard boosts; in particular, the cohomology class of Wigner rotations is invariant under the gauge transformations (\ref{ss17}). So group cohomology provides an answer to the question raised above, namely whether the gauge-dependence of Wigner rotations allows us to remove them altogether: as long as $\cS$ is not the identity, the combination (\ref{s20}) automatically has a non-trivial cohomology class and cannot be removed from the transformation law (\ref{s19}). Since gauge transformations (\ref{s21}) change Wigner rotations only by a coboundary, gauge-invariant observables are quantities that only depend on that cohomology class. In section \ref{sec5} we shall argue that Thomas precession provides such gauge-invariant observables.

%%%%%%%%%%%%%%%%%%%%%%%%%%%%%%%%%%%%%
\subsection{Example: Poincar\'e and Bargmann}
\label{sec3}
%%%%%%%%%%%%%%%%%%%%%%%%%%%%%%%%%%%%%

Here we illustrate the construction of the previous pages with the Poincar\'e and Bargmann groups. We focus on massive representations and work in arbitrary space-time dimension $D+1$. We refer again to \cite[chap.\ 4]{Oblak:2016eij} for a gentler introduction to these matters.

\paragraph*{Poincar\'e.} The (connected) Poincar\'e group is the isometry group of $(D+1)$-dimensional Minkowski space-time; it is the semi-direct product of the (proper, orthochronous) Lorentz group $\text{SO}(D,1)^{\uparrow}$ with the group $\RR^{D+1}$ of space-time translations:
\be
\text{ISO}(D,1)^{\uparrow}
=
\text{SO}(D,1)^{\uparrow}\ltimes\RR^{D+1}.
\nn
\ee
Since this is a semi-direct product, its irreducible unitary representations are given by the Wigner-Mackey method and are classified by orbits of energy-momenta under Lorentz transformations. Each orbit consists of vectors $q$ such that $q_{\mu}q^{\mu}=-M^2$ for some mass squared $M^2$. We focus on a massive particle with positive energy, for which every energy-momentum $q$ on the orbit is a column vector
\be
q=\begin{pmatrix} \sqrt{M^2+\bbq^2}\, \\ \bbq \end{pmatrix}
\nn
\ee
where $\bbq\in\RR^D$ is arbitrary. In the rest frame, the energy-momentum of the particle is just $p=(M,{\textbf 0})^t$. The corresponding little group is $\text{SO}(D)$, consisting of spatial rotations, and the orbit is a hyperbolic space $\text{SO}(D,1)^{\uparrow}/\text{SO}(D)$ diffeomorphic to $\RR^D$. We choose a (generally projective) representation $\cS$ of $\text{SO}(D)$ to specify the particle's spin. At this point, the only ingredient still lacking for (\ref{s19}) is a family of standard boosts. A convenient choice mapping a particle at rest on a particle with momentum $\bbq$ is
\be
g_q
=
\begin{pmatrix}\sqrt{1+\bbq^2/M^2} & \bbq^t/M \\[.3cm]
\bbq/M & \II+\big(\sqrt{1+\bbq^2/M^2}-1\big)\frac{\bbq\bbq^t}{\bbq^2}\end{pmatrix}.
\label{s31}
\ee
Here it is understood that $\bbq$ is a column vector so that $\bbq\bbq^t$ is a symmetric $D\times D$ matrix (with entries $q_iq_j$), while $\bbq^2=\bbq^t\bbq$ and $\II$ is the $D$-dimensional identity matrix.\\

The data just described is enough to explicitly write down the Wigner-Mackey representation (\ref{s19}) for the states of a massive relativistic particle. In keeping with the subject of this paper, we will not describe any of this in detail, except for the Wigner rotations (\ref{s20}). Specifically, let the group element $f$ coincide with a standard boost $g_k$; using (\ref{s31}) and letting $\bx\equiv\bbq/M$, $\by\equiv\bk/M$, one can show (after a lengthy but straightforward calculation) that the Wigner rotation $\sfW_q[g_k]$ is given by
\be
g_q^{-1}g_kg_{g_k^{-1}\cdot q}
=
\begin{pmatrix}
1 & {\textbf 0}^t\\[.3cm]
{\textbf 0} & \II-\frac{\ds\bx\by^t-\by\bx^t+\big(\sqrt{1+\bx^2}\sqrt{1+\by^2}-1\big)\Big(\frac{\bx\bx^t}{\bx^2}+\frac{\by\by^t}{\by^2}-2\frac{\bx(\bx\cdot\by)\by^t}{\bx^2\by^2}\Big)}{\ds\sqrt{1+\bx^2}\sqrt{1+\by^2}-\bx\cdot\by+1}
\end{pmatrix}.
\label{s33}
\ee
This is, as it should, a rotation matrix. In the language of Lorentz symmetry, it embodies the fact that the composition of non-collinear pure boosts does not results in yet another pure boost, but rather contains an extra rotation. For a particle with spin $\cS$, this rotation becomes an operator $\sfW_q[g_k]=\cS[g_q^{-1}g_kg_{g_k^{-1}\cdot q}]$ acting on the spin space. This geometric phenomenon is ultimately responsible for Thomas precession, as we shall see in sections \ref{sec5} and \ref{sec6}. For further details on the computation of Wigner rotations, we refer \eg to \cite{Ferraro:1999eu}.\footnote{Wigner rotations also exist for {\it massless} spinning particles: see \eg \cite{Alsing:2009py}.}

\paragraph*{Bargmann.} We now describe the non-relativistic counterpart of the previous example. In $D$ spatial dimensions, the (connected) Galilei group is a nested semi-direct product
\be
\Gamma(D)
\equiv
\big(\text{SO}(D)\ltimes\RR^D\big)\ltimes\big(\RR^D\times\RR\big)
\label{s35}
\ee
whose elements are quadruples $(f,\bv,\balpha,t)$ where $f\in\text{O}(D)$ is a rotation matrix, $\bv\in\RR^D$ is a velocity vector representing a boost, and $(\balpha,t)\in\RR^D\times\RR$ is a space-time translation. The group operation is
\be
(f,\bv,\balpha,s)\cdot(g,\bw,\bbeta,t)
=
\big(f\cdot g,\bv+f\cdot\bw,\balpha+f\cdot\bbeta+\bv t, s+t\big)
\label{ss35}
\ee
where $f\cdot g$ denotes matrix multiplication while $f\cdot\bw$ and $f\cdot\bbeta$ involve the action of a matrix on a column vector. In contrast to Poincar\'e, space and time, as well as rotations and boosts, live on very different footings. To describe massive particles one needs to add a central extension to (\ref{s35}); thus the Bargmann group is $\widehat{\Gamma}{}(D)=\Gamma(D)\times\RR$, whose elements are $5$-tuples $(f,\bv,\balpha,t,\lambda)$ with $\lambda\in\RR$, subject to the group operation
\be
(f,\bv,\balpha,s,\lambda)\cdot(g,\bw,\bbeta,t,\mu)
=
\Big((f,\bv,\balpha,s)\cdot(g,\bw,\bbeta,t),\lambda+\mu+\bv\cdot f\cdot \bbeta+\frac{\bv^2t}{2}\Big)
\nn
\ee
where the first entry on the right-hand side is a quadruple given by (\ref{ss35}) while $\bv\cdot\bbeta=v^i\beta^i$ denotes the Euclidean scalar product and $\bv^2=v^iv^i$.\\

Just as Poincar\'e, Bargmann is a semi-direct product with an Abelian normal subgroup of space-time translations, so its irreducible unitary representations are specified by orbits of energy-momentum-mass vectors under rotations and boosts. The classification is somewhat harder than in the relativistic case due to the intricate group structure; see \eg \cite{Inonu} or \cite[sec.\ 4.4.2]{Oblak:2016eij}. Here we focus on massive particles, for which the spatial momentum $\bbq$ and mass $M$ determine the energy according to $E=\bbq^2/2M$. In the rest frame, $\bbq={\textbf 0}$. The corresponding little group $\text{SO}(D)$ consists of rotations and the orbit is $(\text{SO}(D)\ltimes\RR^D)/\text{SO}(D)\cong\RR^D$; spin is a representation $\cS$ of $\text{SO}(D)$. As for boosts, they are much simpler than in the relativistic case: to map a particle at rest on a particle with momentum $\bbq$, just apply a boost with velocity $\bv=\bbq/M$:
\be
g_{\bbq}
=
(\II,\bbq/M)\in\text{SO}(D)\ltimes\RR^D.
\label{s37}
\ee
At this point we have all the information needed to write the non-relativistic version of the Wigner-Mackey formula (\ref{s19}). As before we only focus on Wigner rotations. In the present case we let $f\in\text{SO}(D)$ be an arbitrary rotation, $\bv\in\RR^D$ an arbitrary boost, and investigate the corresponding combination (\ref{s20}) at momentum $\bbq$:
\be
g_{\bbq}^{-1}\cdot(f,\bv)\cdot g_{(f,\bv)^{-1}\cdot\bbq}
=g_{\bbq}^{-1}\cdot(f,\bv)\cdot g_{f^{-1}\cdot\bbq-Mf^{-1}\cdot\bv}
\refeq{ss35}
(f,{\textbf 0}).
\nn
\ee
The last equality is a remarkable fact: it says that Wigner rotations project $(f,\bv)$ on the rotation $f$ alone, for any momentum of a non-relativistic particle. It is a radically different conclusion than the one of the relativistic case. In particular, the non-relativistic Wigner rotations associated with pure boosts always vanish,
\be
g_{\bbq}^{-1}\cdot g_{\bk}\cdot g_{g_{\bk}^{-1}\cdot\bbq}
=
(\II,{\textbf 0}),
\label{s41}
\ee
in contrast to the non-trivial relativistic rotation (\ref{s33}). In sections \ref{sec5} and \ref{sec6} we shall relate these statements to Thomas precession and Berry phases on momentum orbits.

%%%%%%%%%%%%%%%%%%%%%%%%%%%%%%%%%%%%%%%%%%%%%%%%%%%%%%%%%%%
\section{Infinitesimal Wigner rotations}
\label{sec4}
%%%%%%%%%%%%%%%%%%%%%%%%%%%%%%%%%%%%%%%%%%%%%%%%%%%%%%%%%%%

As a preliminary step towards Berry phases and Thomas precession, in this section we differentiate the Wigner-Mackey formula (\ref{s19}) to obtain unitary representations of the Lie algebra of $G\ltimes A$. Surprisingly, we were unable to find this computation in the literature. For our purposes, the most important result will be a Lie-algebraic analogue of the Wigner rotation (\ref{s20}), which we will study in detail.

%%%%%%%%%%%%%%%%%%%%%%%%%%%%%%%%%%%%%
\subsection{Differentiating Wigner-Mackey}
\label{sec41}
%%%%%%%%%%%%%%%%%%%%%%%%%%%%%%%%%%%%%

The Lie algebra of $G\ltimes A$ is a semi-direct sum $\mg\inplus A$ where $\mg$ is the Lie algebra of $G$, and its elements are pairs $(X,\alpha)$ where $X\in\mg$ and $\alpha\in A$.\footnote{Since $A$ is a vector group, its Lie algebra is the vector space $A$ endowed with a trivial Lie bracket.} We wish to differentiate the group representation (\ref{s19}); explicitly, for any $(X,\alpha)\in\mg\inplus A$ we define
\be
\cu[(X,\alpha)]
\equiv
\left.\frac{d}{dt}\right|_{t=0}\cU[(e^{tX},t\alpha)]
\label{ss43}
\ee
where $e^X\in G$ is the exponential of $X\in\mg$. Since $\cU$ is unitary, $\cu[(X,\alpha)]$ is an anti-Hermitian operator acting on the Hilbert space of $\mh$-valued wavefunctions on $\cO_p$. To obtain $\cu$ we need to differentiate eq.\ (\ref{s19}) with respect to $f$ and $\alpha$; we do this by treating one by one the three terms on the right-hand side of that formula.

\paragraph{Scalar contribution.} First, the differential of the exponential term is just $\der_t|_{0}e^{i\bra q,t\alpha\ket}=i\bra q,\alpha\ket$. Secondly, let us take $f=e^{tX}$ and differentiate the term $\Psi(f^{-1}\cdot q)$:
\be
\left.\frac{d}{dt}\right|_{0}\Psi(e^{-tX}\cdot q)
=
\text{d}\Psi_q\Big(
\left.\frac{d}{dt}\right|_{0}(e^{-tX}\cdot q)
\Big).
\label{s43}
\ee
Here we are assuming that $\Psi$ is differentiable and write $\text{d}\Psi_q$ for its differential (pushforward) at $q\in\cO_p$.\footnote{The set of smooth functions is dense in $L^2(\cO_p)$, so this assumption entails no loss of generality.} The argument of $\text{d}\Psi_q$ is a vector tangent to $\cO_p$ at $q$. In fact, it is the {\it fundamental vector field} $\xi_X$ generating the action of $G$ on $\cO_p$, evaluated at $q$:\footnote{We are defining $\xi_X$ with an exponential path $e^{tX}$, but in fact any curve $\gamma(t)$ in $G$ such that $\gamma(0)=e$ and $\dot\gamma(0)=X$ would do the job; this is important for some of the computations below.}
\be
(\xi_X)_q
\equiv
\left.\frac{d}{dt}\right|_{t=0}\big(e^{tX}\cdot q\big).
\label{ss44}
\ee
With this notation we can rewrite (\ref{s43}) as $\der_t|_{0}\Psi(e^{-tX}\cdot q)=-\text{d}\Psi_q(\xi_X)_q=-\big(\xi_X\cdot\Psi\big)(q)$. All in all, the contribution to (\ref{ss43}) of the spin-independent terms of (\ref{s19}) is
\be
\left.\frac{d}{dt}\right|_{0}
\Big[
e^{i\bra q,t\alpha\ket}\Psi(e^{-tX}\cdot q)
\Big]
=
i\bra q,\alpha\ket\Psi(q)-\big(\xi_X\cdot\Psi\big)(q).
\label{t44}
\ee
In a scalar representation (no spin), this would be the end of the story.\\

At this point a few words are in order regarding the fundamental vector field (\ref{ss44}). For generic $X\in\mg$ and $q\in\cO_p$, the vector $(\xi_X)_q$ does not vanish because $e^{tX}$ does not leave $q$ fixed. However, by construction, any point on the orbit has a non-trivial stabilizer, which for $p$ is the little group $G_p$. Accordingly, from now on we refer to the Lie algebra of the little group as the {\it little algebra}, denoted $\mg_p$; it consists of those Lie algebra elements $X\in\mg$ such that $(\xi_X)_p=0$.

\paragraph{Wigner generators.} To complete the computation of (\ref{ss43}), it remains to differentiate the Wigner rotation (\ref{s20}). Letting $f=e^{tX}$ as before, we have
\be
\left.\frac{d}{dt}\right|_{0}
\cS\big[g_q^{-1}e^{tX}g_{e^{-tX}\cdot q}\big]
=
\cs\bigg[\left.\frac{d}{dt}\right|_{0}\Big(g_q^{-1}e^{tX}g_q\Big)
+\left.\frac{d}{dt}\right|_{0}\Big(g_q^{-1}g_{e^{-tX}\cdot q}\Big)\bigg]
\label{s45}
\ee
where $\cs$ is the representation of the little algebra obtained by differentiating $\cS$. Note that the two terms in the argument of $\cs$ do {\it not} separately belong to the little algebra, so one cannot split the right-hand side of (\ref{s45}) as a sum of two operators $\cs[...]$. The first term is
\be
\left.\frac{d}{dt}\right|_{0}g_q^{-1}e^{tX}g_q
=
\Ad_{g_q^{-1}}X
\label{ss45}
\ee
where $\Ad$ is the adjoint representation of $G$. To deal with the second term, recall that the (left) {\it Maurer-Cartan form} on $G$ is defined as the $\mg$-valued one-form $\Theta$ given by
\be
\Theta_f
\equiv
\text{d}\big(L_{f^{-1}}\big)_f
\qquad\forall f\in G,
\label{ss46}
\ee
where $L_{f^{-1}}$ denotes left multiplication by $f^{-1}$ and $\text{d}(L_{f^{-1}})$ is its differential (pushforward). When $G$ is a matrix group, the matrix entries define local coordinates on $G$ and the Maurer-Cartan form is typically written as $\Theta=f^{-1}\text{d}f$; but for the sake of generality, let us use the abstract definition (\ref{ss46}). In these terms the second piece of (\ref{s45}) is
\be
\left.\frac{d}{dt}\right|_{0}\Big(g_q^{-1}g_{e^{-tX}\cdot q}\Big)
=
-\Theta_{g_q}\text{d}g_q(\xi_X)_q
\label{s46}
\ee
where $\xi_X$ is the fundamental vector field (\ref{ss44}) and $\text{d}g_q$ is the differential at $q\in\cO_p$ of the family of standard boosts (\ref{s17b}), which we assume to be smooth. In fact, eq.\ (\ref{s46}) contains the pullback of the Maurer-Cartan form (\ref{ss46}) by these standard boosts: $\Theta_{g_q}\circ\text{d}g_q=(g^*\Theta)_q$. (This is often written as a pure gauge field configuration $g_q^{-1}\text{d}g_q$.) With this notation we can combine eqs.\ (\ref{s46}) and (\ref{ss45}) to write (\ref{s45}) as
\be
\left.\frac{d}{dt}\right|_{0}\sfW_q[e^{tX}]
=
\cs\big[\Ad_{g_q^{-1}}X-(g^*\Theta)_q(\xi_X)_q\big]
\equiv
\sfw_q[X],
\label{s47}
\ee
where we stress once more that the two terms in the argument of $\cs$ do not separately belong to the little algebra, though their combination does. From now on we refer to $\sfw_q[X]$ as an infinitesimal Wigner rotation, or {\it Wigner generator} for short.\\

By now we can evaluate the Lie algebra representation (\ref{ss43}). Putting together (\ref{t44}) and (\ref{s47}), the differential of the transformation law (\ref{s19}) at the identity is
\be
\big(\cu[(X,\alpha)]\cdot\Psi\big)(q)
=
\Big(i\langle q,\alpha\rangle+\cs\big[\Ad_{g_q^{-1}}X-(g^*\Theta)_q(\xi_X)_q\big]\Big)\Psi(q)
-(\xi_X\Psi)(q).
\nn
\ee
To rewrite this more compactly, we remove the argument $q$ so that
\be
\boxed{\cu[(X,\alpha)]\cdot\Psi
=
\Big(i\langle\cdot,\alpha\rangle+\cs\big[\Ad_{g^{-1}}X-g^*\Theta(\xi_X)\big]-\xi_X\Big)\Psi}
\label{ss47}
\ee
where $\langle\cdot,\alpha\rangle$ is the real function on $\cO_p$ that maps $q$ on the number $\langle q,\alpha\rangle$.\\

Formula (\ref{ss47}) is our first key result. It says that infinitesimal translations act by multiplication on wavefunctions in momentum space, while rotations or boosts act as translations on a momentum orbit (due to the differential operator $\xi_X$) and rotate wavefunctions in their internal (spin) space with the Wigner generator (\ref{s47}); the latter will be studied in detail in the upcoming pages. To the best of our knowledge, eq.\ (\ref{ss47}) does not appear in the literature, though there should be at least one other way to derive it. Indeed, most Wigner-Mackey representations can be obtained by geometric quantization of the coadjoint orbits of $G\ltimes A$. (See \cite{Rawnsley1975,Li1993} for mathematical aspects and \cite{Robson:1994zq} for examples and further references.) For example, the scalar transformation law (\ref{t44}) directly follows from the action of $\mg\inplus A$ on polarized sections on $T^*\cO_p$ (see \eg \cite[sec.\ 5.4.4]{Oblak:2016eij}). It should be possible to similarly prove the more general formula (\ref{ss47}) from geometric quantization, but we are not aware of any reference that exhibits this computation.

%%%%%%%%%%%%%%%%%%%%%%%%%%%%%%%%%%%%%
\subsection{Comments on Wigner generators}
\label{sec42}
%%%%%%%%%%%%%%%%%%%%%%%%%%%%%%%%%%%%%

Until the end of this section we focus on the operator (\ref{s47}); in particular our goal is to investigate the Lie-algebraic version of the properties listed in section \ref{sec23}.

\paragraph{Projectors.} Consider the argument of $\cs$ in (\ref{s47}), which is an element of the little algebra:
\be
\Ad_{g_q^{-1}}X-(g^*\Theta)_q(\xi_X)_q\,\in\,\mg_p
\label{s61}
\ee
 As mentioned before, the two terms of this expression do not separately belong to $\mg_p$ and the role of the second term in (\ref{s61}) is to set to zero all the components of $\Ad_{g_q^{-1}}X$ that are not along $\mg_p$. In fact, the map sending $\Ad_{g_q^{-1}}X$ on (\ref{s61}) is a projection:

\paragraph{Lemma.} Let $\big\{g_q\big|q\in\cO_p\big\}$ be a set of standard boosts. Then, for any $q\in\cO_p$, the map
\be
\pi_q:\mg\rightarrow\mg:X\mapsto X-(g^*\Theta)_q\big(\xi_{\Ad_{g_q}X}\big)_q
\label{ss13b}
\ee
is a linear projection operator in the sense that $(\pi_q)^2=\pi_q$.

\paragraph{Proof.} Let $X\in\mg$; let $q\in\cO_p$ and $Y\equiv\Ad_{g_q}X$. One then finds by brute force that
\be
\pi_q^2(X)
=
\pi_q\Big(X-(g^*\Theta)_q\big(\xi_Y\big)_q\Big)
=
\pi_q(X)
-
(g^*\Theta)_q\Big[(\xi_Y)_q-\big(\xi_{\Ad_{g_q}(g^*\Theta)_q(\xi_Y)_q}\big)_q\Big].
\label{s13b}
\ee
In the very last term we have $\Ad_{g_q}(g^*\Theta)_q(\xi_Y)_q=\der_t|_{0}\big[g_{e^{tY}\cdot q}g_q^{-1}\big]$, so that
\be
\big(\xi_{\Ad_{g_q}(g^*\Theta)_q(\xi_Y)_q}\big)_q
=
\big(\xi_{\left.\frac{d}{dt}\right|_{0}(g_{e^{tY}\cdot q}g_q^{-1})}\big)_q
\refeq{ss44}
\left.\frac{d}{dt}\right|_{0}\Big[g_{e^{tY}\cdot q}g_q^{-1}\cdot q\Big]
=
\left.\frac{d}{dt}\right|_{0}\Big[g_{e^{tY}\cdot q}\cdot p\Big]
\refeq{s17}
(\xi_Y)_q.
\nn
\ee
Using (\ref{s13b}) this implies that $\pi_q^2(X)=\pi_q(X)-(g^*\Theta)_q\big[(\xi_Y)_q-(\xi_Y)_q\big]=\pi_q(X)$. Furthermore the definition (\ref{ss13b}) ensures that $\pi_q$ is linear, so it is indeed a projector.\hfill$\blacksquare$\\

This lemma demystifies the seemingly awkward combination of terms in (\ref{s47}) and (\ref{s61}): it allows us to write Wigner generators as
\be
\sfw_q[X]
=
\cs\big[\pi_q\big(\Ad_{g_q}X\big)\big],
\nn
\ee
\ie as operators in spin space obtained by projecting $\Ad_{g_q}X$ to the little algebra in a momentum-dependent way. Note that in general, the projection (\ref{ss13b}) genuinely depends on momentum in the sense that $\pi_q\neq\pi_k$ when $q\neq k$; this can be verified by fixing $X\in\mg$ and evaluating the differential of the map $\cO_p\rightarrow\mg:q\mapsto\pi_q(X)$.

\paragraph{Gauge invariance and cohomology.} We now study the gauge-theoretic properties of Wigner generators. Consider a change of standard boosts (`gauge transformation') as in (\ref{ss17}); how does it affect (\ref{s47})? To answer this we return to (\ref{s45}) and find
\be
\left.\frac{d}{dt}\right|_{0}
\cS\Big[h_q^{-1}g_q^{-1}e^{tX}g_{e^{-tX}\cdot q}h_{e^{-tX}\cdot q}\Big]
=
\sfw_q[\Ad_{h_q^{-1}}X]-\cs[\theta_{h_q}](\text{d}h)_q(\xi_X)_q
\nn
\ee
where $\theta$ is the Maurer-Cartan form of the little group $G_p$. We can rewrite this more compactly by recognizing $\cs[\theta]\text{d}h$ as the pullback of $\cs[\theta]$ by the section $h:\cO_p\rightarrow G_p:q\mapsto h_q$; removing the argument $q$, we conclude that under gauge transformations (\ref{ss17}) the infinitesimal Wigner rotation (\ref{s47}) changes as $\sfw\mapsto\widetilde{\sfw}$, with
\be
\widetilde{\sfw}{}[X]
=
\cS[h]^{-1}\sfw[X]\cS[h]-\cs[h^*\theta]\xi_X.
\label{s53}
\ee
This is the Lie-algebraic version of eq.\ (\ref{s21}). It confirms that Wigner generators are not gauge-invariant, though their transformation law is somewhat similar to that of a gauge field. We will return to this in section \ref{sec52}, where we shall build a Berry gauge connection based on the Wigner generator (\ref{s47}).\\

Just as their group-theoretic cousins, the Wigner generators (\ref{s47}) have interesting cohomological properties. Namely, for each $X\in\mg$ one can think of $\sfw[X]$ as a map that sends $q\in\cO_p$ on the operator $\sfw_q[X]\in\text{End}(\mh)$ given by (\ref{s47}). This provides a map
\be
\sfw:\mg\rightarrow C^{\infty}\big(\cO_p,\text{End}(\mh)\big):X\mapsto\sfw[X]
\label{ss55}
\ee
which is the Lie-algebraic counterpart of (\ref{s23b}). Using the cocycle property (\ref{s23}), one finds that Wigner generators are compatible with the Lie bracket of $\mg$ in the sense that
\be
\sfw\big[[X,Y]\big]
=
\big[\sfw[X],\sfw[Y]\big]-\xi_X\cdot\sfw[Y]+\xi_Y\cdot\sfw[X].
\label{s55}
\ee
Here the bracket on the right-hand side is the commutator in $\text{End}(\mh)$, while $\xi_X\cdot\sfw[Y]$ denotes the action of the vector field $\xi_X$ on the $\text{End}(\mh)$-valued function $\sfw[Y](q)=\sfw_q[Y]$. Eq.\ (\ref{s55}) says that the map (\ref{ss55}) is a one-cocycle on $\mg$ taking its values in a space of operator-valued functions.\footnote{More abstractly, Wigner generators provide a representation of the action Lie algebroid $\mg\ltimes\cO_p$.} One can again verify that the representation (\ref{ss47}) is equivalent to a scalar one if and only if the corresponding Wigner generators are cohomologically trivial, \ie if $\sfw_q[X]=-\Omega(q)^{-1}\cdot\big(\xi_X\cdot\Omega\big)(q)$ for some function $\Omega:\cO_p\rightarrow\text{GL}(\mh)$. In that language, the Wigner-Mackey theorem implies that, by construction, the operators (\ref{s47}) define a non-trivial cocycle whenever the spin representation $\cs$ is non-trivial.

%%%%%%%%%%%%%%%%%%%%%%%%%%%%%%%%%%%%%
\subsection{Splitting Wigner generators}
\label{sec43}
%%%%%%%%%%%%%%%%%%%%%%%%%%%%%%%%%%%%%

So far our observations were independent of the choice of standard boosts on the orbit: all our statements were covariant under the `gauge transformations' (\ref{ss17}). But now we shall impose a specific partial gauge-fixing condition in order to uncover further properties of Wigner rotations. The motivation stems from the sum of terms in (\ref{s61}): the first term depends directly on $X$, whereas the second only depends on it through the vector field $\xi_X$. From a gauge-theoretic perspective it is tempting to split that sum in two pieces, each belonging to the little algebra, with one piece depending on $X\in\mg$ only through $\xi_X$. To perform that splitting, let us require that standard boosts reduce to the identity at $p$:
\be
g_p=e.
\label{s59}
\ee
This condition is satisfied by the standard boosts (\ref{s31}) and (\ref{s37}) chosen above for Poincar\'e and Bargmann, respectively. Any other family of standard boosts can be brought into such a form with a gauge transformation (\ref{ss17}): it suffices to choose $h_p=g_p^{-1}$. When (\ref{s59}) holds, the standard boosts define a section (\ref{s17b}) whose differential at $p$ is a linear map $\text{d}g_p:T_p\cO_p\rightarrow\mg$. Keeping in mind the Lie algebra element (\ref{s61}), we would like to act with this map on a vector $(\xi_X)_q$ at $q$; since the differential is taken at $p$, we first translate $(\xi_X)_q$ to $p$ using the action (\ref{s15}) of $G$ on $\cO_p$, which gives a vector
\be
\text{d}(g_q^{-1}\cdot)_q(\xi_X)_q
=
\big(\xi_{\Ad_{g_q^{-1}}X}\big)_p\,\in\,T_p\cO_p
\label{ss63}
\ee
where, for any $f\in G$, the notation $\text{d}(f\cdot)_k$ means `the differential at $k$ of the map $q\mapsto f\cdot q$.' (To prove the equality in (\ref{ss63}) we used the definition (\ref{ss44}) of $\xi_X$.) The key observation now is that $\text{d}g_p$ acting on (\ref{ss63}) produces a Lie algebra element that can be combined with the two pieces of (\ref{s61}) in such a way that they {\it separately} belong to the little algebra:

\paragraph{Lemma.} The Lie algebra elements
\be
\Ad_{g_q^{-1}}X-\text{d}g_p\text{d}(g_q^{-1}\cdot)_q(\xi_X)_q
\qquad\text{and}\qquad
\Big((g^*\Theta)_q-\text{d}g_p\text{d}(g_q^{-1}\cdot)_q\Big)(\xi_X)_q
\label{t63}
\ee
both belong to the little algebra $\mg_p$, for any $X\in\mg$ and any $q\in\cO_p$.

\paragraph{Proof.} The difference of the two expressions in (\ref{t63}) coincides with (\ref{s61}), which belongs to the little algebra. As the latter is a vector space, if we prove that one of the two quantities in (\ref{t63}) belongs to $\mg_p$, then so does the other. Accordingly, it suffices to prove that the first expression in (\ref{t63}) belongs to the little algebra, \ie that the corresponding fundamental vector field (\ref{ss44}) vanishes at $p$. To see this we first compute
\be
\big(\xi_{\Ad_{g_q^{-1}}X}\big)_p
\refeq{ss44}
\left.\frac{d}{dt}\right|_{0}\big(\exp[t\Ad_{g_q^{-1}}X]\cdot p\big)
=
\left.\frac{d}{dt}\right|_{0}\big(g_q^{-1}e^{tX}g_q\cdot p\big)
=
\left.\frac{d}{dt}\right|_{0}\big(g_q^{-1}e^{tX}\cdot q\big).
\label{s64}
\ee
On the other hand, using (\ref{ss63}) one finds
\be
\Big(\xi_{\ds\text{d}g_p\text{d}(g_q^{-1}\cdot)_q(\xi_X)_q}\Big)_p
=\left.\frac{d}{dt}\right|_{0}\Big[\exp\big[t\text{d}g_p(\xi_{\Ad_{g_q^{-1}}X})_p\big]\cdot p\Big]
=\left.\frac{d}{dt}\right|_{0}\big[g_{\exp[t\Ad_{g_q^{-1}}X]\cdot p}\cdot p\big]
\nn
\ee
which coincides with (\ref{s64}). Thus the fundamental vector field associated with the first Lie algebra element in (\ref{t63}) vanishes at $p$, as was to be proved.\hfill$\blacksquare$\\

This lemma allows us to split infinitesimal Wigner rotations (\ref{s47}) as
\be
\sfw_q[X]
=
\cs\big[\Ad_{g_q^{-1}}X-\text{d}g_p\text{d}(g_q^{-1}\cdot)_q(\xi_X)_q\big]
-\cs\big[(g^*\Theta)_q(\xi_X)_q-\text{d}g_p\text{d}(g_q^{-1}\cdot)_q(\xi_X)_q\big]
\nn
\ee
where the right-hand side is well-defined since both arguments of $\cs$ belong to the little algebra. The last term, in particular, only depends on $X$ through the vector field $\xi_X$, so it is tempting to interpret it as arising from a $\mg_p$-valued connection one-form
\be
\cA
=
\cs\big[g^*\Theta-\text{d}g_p\text{d}(g^{-1}\cdot)\big].
\label{s67}
\ee
In fact we shall see in section \ref{sec52} that the Berry connection associated with adiabatic changes of reference frames takes precisely the same form. For now, we simply think of (\ref{s67}) as a convenient tool to study Wigner rotations. One may wonder, for instance, if there are situations where (\ref{s67}) vanishes identically; the answer is as follows:

\paragraph{Lemma.} The one-form (\ref{s67}) vanishes if and only if the Wigner rotations (\ref{s20}) associated with standard boosts are trivial, \ie if
\be
\cS[g_q^{-1}\,g_k\,g_{g_k^{-1}\cdot q}]=\II
\qquad
\forall\,k,q\in\cO_p.
\label{s69}
\ee

\paragraph{Proof.} If $\cS$ is trivial, then (\ref{s69}) certainly holds and (\ref{s67}) vanishes; so let us focus on the more interesting case where $\cS$ is a non-trivial representation of $G_p$. The trick will be to rewrite (\ref{s67}) in a way that explicitly relates it to a derivative of $\cS$. Let $\gamma(t)$ be a path on the orbit $\cO_p$; it defines a tangent vector $\dot\gamma(t)$ that we can pair with (\ref{s67}):
\begin{align}
\cs\big[(g^*\Theta)_{\gamma(t)}\dot\gamma(t)-\text{d}g_p\text{d}(g_{\gamma(t)}^{-1}\cdot)_{\gamma(t)}\dot\gamma(t)\big]
&\refeq{ss46} \cs\Big[\left.\frac{d}{d\tau}\right|_{\tau=t}\Big(g_{\gamma(t)}^{-1}g_{\gamma(\tau)}\Big)
-\left.\frac{d}{d\tau}\right|_{\tau=t}\Big(g_{g_{\gamma(t)}^{-1}\cdot\gamma(\tau)}\Big)\Big]\nn\\
\label{s71}
&\stackrel{\text{\textcolor{white}{(3.7)}}}{=}-\left.\frac{d}{d\tau}\right|_{\tau=t}\cS\Big[g_{\gamma(\tau)}^{-1}g_{\gamma(t)}g_{g_{\gamma(t)}^{-1}\cdot\gamma(\tau)}\Big].
\end{align}
The last line is the derivative of a {\it finite} (as opposed to infinitesimal) Wigner rotation, which allows us to relate the vanishing of (\ref{s67}) to the triviality (\ref{s69}) of Wigner rotations. Indeed, saying that (\ref{s67}) vanishes means that it gives zero when paired with any tangent vector on $\cO_p$, \ie it is equivalent to the vanishing of the time derivative (\ref{s71}) for any choice of path $\gamma$. Since $\gamma$ is arbitrary, (\ref{s71}) vanishes if and only if the operator $\cS[g_q^{-1}g_kg_{g_k^{-1}\cdot q}]$ is independent of $k$ and $q$, \ie if it takes a constant value on the orbit. But standard boosts are continuous by assumption and the gauge condition (\ref{s59}) requires $g_p=e$, so the vanishing of (\ref{s71}) implies eq.\ (\ref{s69}). Conversely, if Wigner rotations are trivial as in (\ref{s69}), then eq.\ (\ref{s71}) ensures that (\ref{s67}) vanishes.\hfill$\blacksquare$\\

Note that Wigner rotations of standard boosts are precisely those we evaluated in section \ref{sec3} for Poincar\'e and Bargmann: in the former case we found in (\ref{s33}) that these rotations are non-trivial, while in the latter we saw in (\ref{s41}) that they always vanish.

%%%%%%%%%%%%%%%%%%%%%%%%%%%%%%%%%%%%%%%%%%%%%%%%%%%%%%%%%%%
\section{Thomas precession as a Berry phase}
\label{sec5}
%%%%%%%%%%%%%%%%%%%%%%%%%%%%%%%%%%%%%%%%%%%%%%%%%%%%%%%%%%%

In this section we study the main observable consequence of Wigner rotations --- Thomas precession ---, which we describe as a Berry phase associated with Wigner-Mackey representations. Accordingly, we start by recalling in general terms how unitary group representations lead to Berry phases, before applying that approach to semi-direct products.\\

Incidentally, the literature already contains many references that treat Thomas precession as a holonomy \cite{Mathur:1991ynd,Aravind}, or equivalently a Berry phase \cite{Brezov:2015aja,Stone:2014fja}. However, it seems that none of them use the Wigner-Mackey description of one-particle states; instead, most focus on the special case of Poincar\'e symmetry and rely on spin-specific tools such as the Dirac equation. Our approach, by contrast, will not only hold for any spin, but will in fact allow us to describe Thomas precession for any symmetry group with a semi-direct product structure. In section \ref{sec6} we will apply this method to the Poincar\'e and Bargmann groups; the BMS group will be treated in a separate paper \cite{OblakThomas}.

%%%%%%%%%%%%%%%%%%%%%%%%%%%%%%%%%%%%%
\subsection{Berry phases in group representations}
\label{sec51}
%%%%%%%%%%%%%%%%%%%%%%%%%%%%%%%%%%%%%

As a preparation for Thomas precession, our goal here is to exhibit certain Berry holonomies that appear in unitary representations of Lie groups \cite{Berry:1984jv,jordan1988berry}. In order to leave room for non-trivial spin spaces, we will deal with generally degenerate eigenvalues of the Hamiltonian. This will require the non-Abelian generalization of Berry phases first described in \cite{Wilczek:1984dh}; the non-degenerate, Abelian version of the argument can be found \eg in \cite{Oblak:2017ect}. In contrast to the rest of this paper, in this section we use the Dirac notation.

\paragraph{Berry phases.} Consider a (connected, simply connected) Lie group $G$ and a unitary representation $\cU$ thereof. Think of $G$ as a symmetry group consisting of `changes of reference frames' and assume that it contains a one-parameter subgroup of transformations that can be interpreted as time translations. Each such transformation corresponds to a group element $e^{tX_0}$ for some fixed $X_0$ belonging to the Lie algebra $\mg$ of $G$. From that perspective, $X_0\in\mg$ is the generator of time translations and the evolution operator is
\be
\cU[e^{tX_0}]
=
e^{t\cu[X_0]}
\label{s80}
\ee
where $\cu$ is the Lie algebra representation corresponding to $\cU$ by differentiation. This is to say that the Hamiltonian is $H=i\cu[X_0]$. Crucially however, the latter statement relies on an arbitrary choice of reference frame: if $f\in G$ relates two observers $A$ and $B$, and if $A$ sees an evolution operator (\ref{s80}), then $B$ will observe a generally different one,
\be
\cU[f]\cU[e^{tX_0}]\cU[f]^{-1}
=
\cU[e^{t\Ad_fX_0}]
=
e^{t\cu[\Ad_fX_0]},
\label{s81}
\ee
corresponding to a different Hamiltonian $H'=i\cu[\Ad_fX_0]=\cU[f]H\cU[f]^{-1}$. Thus, given the representation $\cU$, one obtains a family of Hamiltonian operators labelled by $f\in G$. One can then think of $G$ as a space of parameters whose adiabatic variations generally lead to geometric phases picked along time evolution by any wavefunction. In particular, closed paths in parameter space lead to Berry phases \cite{Berry:1984jv}.\\

Concretely, let $E$ be an $N$-fold degenerate eigenvalue of $H=i\cu[X_0]$, and let $|\phi_1\ket,...,|\phi_N\ket$ be normalized, mutually orthogonal eigenvectors of $H$ for this eigenvalue. For definiteness we assume that the latter is isolated, though this assumption will fail to hold for semi-direct products. Now suppose that the system is initially in a state $|\psi(0)\ket=\cU[f(0)]|\phi_i\ket$ for some $f(0)\in G$, and evolves according to the time-dependent Schr\"odinger equation
\be
i\der_t|\psi(t)\ket
=
\cU[f(t)]H\cU[f(t)]^{-1}|\psi(t)\ket
\label{ss83}
\ee
where the path $f(t)$ in $G$ represents time-dependent changes of reference frames. Provided $f(t)$ varies sufficiently slowly, the adiabatic theorem \cite{Born1928,Avron:1998th} ensures that
\be
|\psi(t)\ket
\sim
e^{-iEt}\,\Omega_{ji}(t)\cU[f(t)]|\phi_j\ket
\qquad\text{(in the adiabatic limit)}
\label{s83}
\ee
with an implicit sum over $j=1,...,N$. Here $(\Omega_{ij}(t))$ is a time-dependent unitary $N\times N$ matrix; on account of (\ref{ss83}) it satisfies the differential equation
\be
\der_t\Omega_{ik}(t)\big(\Omega(t)^{-1}\big)_{kj}
=
-\bra\phi_i|\cU[f(t)]^{-1}\der_t\cU[f(t)]|\phi_j\ket.
\label{t84}
\ee
At this point we introduce the (non-Abelian) anti-Hermitian Berry connection
\be
A_f
\equiv
\bra\vec{\phi}|\big(\cU[\cdot]^{-1}\text{d}\cU[\cdot]\big)_f|\vec{\phi}\ket^t
=
\bra\vec{\phi}|\cu[\Theta_f]|\vec{\phi}\ket^t
\label{ss84}
\ee
where $|\vec{\phi}\ket$ is a column vector whose entries are $|\phi_1\ket,...,|\phi_N\ket$, while $\text{d}$ is the exterior derivative on $G$ and $\Theta$ is the Maurer-Cartan form (\ref{ss46}). In these terms eq.\ (\ref{t84}) reads $\der_t\Omega\cdot\Omega^{-1}=-A_f(\dot f)$ and its solution is a Wilson line
\be
\Omega(t)
=
P\exp\Big[-\int_f A\Big].
\label{s87}
\ee
The integral over $f$ is evaluated between the times $0$ and $t$. For closed paths, \ie when $f(T)=f(0)$ for some time $T>0$, the matrix (\ref{s87}) becomes a {\it Berry holonomy}
\be
\Omega(T)
\equiv
B_{\vec\phi}[f]
=
P\exp\Big[-\oint_fA\Big]
\refeq{ss84}
P\exp\Big[-\oint_f\bra\vec{\phi}|\cu[\Theta]|\vec{\phi}\ket^t\Big]
\label{ss87}
\ee
whose eigenvalues are complex numbers with unit norm; their phases are Berry phases.

\paragraph{Remarks.} Various comments are in order regarding formula (\ref{ss87}). As a prerequisite, we define the stabilizer of the states $|\phi_1\ket,...,|\phi_N\ket$:\footnote{This definition fails for semi-direct products --- more on that in section \ref{sec54}.}
\be
G_{\vec\phi}
\equiv
\Big\{
h\in G
\,\Big|\,
\cU[h]|\phi_i\ket=\Lambda_{ij}|\phi_j\ket\;\forall\,i=1,...,N,\;(\Lambda_{ij})\text{ unitary}
\Big\}.
\label{s88}
\ee
It is the subgroup of $G$ whose elements rotate the vectors $|\phi_i\ket$ among themselves. Note that, since $\cU$ is unitary, the matrices $\cS_{ij}[h]\equiv\bra\phi_i|\cU[h]|\phi_j\ket$ provide an $N$-dimensional unitary representation $\cS$ of $G_{\vec\phi}$. One can think of it as an analogue of the spin representation that appears in the Wigner-Mackey construction, which is why we call it $\cS$.\\

Now consider the state vector (\ref{s83}), which solves the Schr\"odinger equation (\ref{ss83}) in the adiabatic limit. In writing that vector we have arbitrarily declared that it is a linear combination of states $\cU[f(t)]|\phi_j\ket$, while we could just as well have used $\cU[f(t)\cdot h(t)]|\phi_j\ket$ for any path $h(t)$ contained in the stabilizer (\ref{s88}). With this different convention, eq.\ (\ref{t84}) governing the time-dependence of $\Omega(t)$ would have been replaced by
\be
\der_t\Omega\cdot\Omega^{-1}
=
-\cS[h(t)]^{-1}\bra\vec\phi|\cU[f(t)]^{-1}\der_t\cU[f(t)]|\vec\phi\ket^t\cS[h(t)]-\cS[h(t)]^{-1}\cdot\der_t\cS[h(t)].
\nn
\ee
This amounts to transforming the Berry connection (\ref{ss84}) as
\be
A_f\mapsto \tilde A_{f\cdot h}
=
\cS[h]^{-1}A_f\cS[h]+\cS[h]^{-1}\cdot\text{d}\cS[h],
\nn
\ee
so the replacement of $f(t)$ by $f(t)\cdot h(t)$ is akin to a gauge transformation with gauge group $G_{\vec\phi}$. From that perspective the connection (\ref{ss84}) is a gauge field valued in the Lie algebra of the stabilizer (more precisely, in the space of operators representing that algebra through $\cS$). As for the Wilson line (\ref{s87}), it is {\it not} invariant under such transformations; this remains true even if the curves $f(t)$ and $h(t)$ are both closed, so the holonomy (\ref{ss87}) is not directly observable, though its eigenvalues (hence their Berry phases) are.\\

From a gauge-theoretic standpoint one may wonder if the holonomies (\ref{ss87}) have any chance of being non-trivial at all. Indeed, the gauge connection (\ref{ss84}) is essentially the Maurer-Cartan form sandwiched between two $|\phi_i\ket$'s. Since the Maurer-Cartan form is `pure gauge' ($\Theta_f=f^{-1}\text{d}f$), it is flat in the sense that for any two vector fields $\xi,\zeta$ on $G$,
\be
\big(\text{d}\Theta\big)(\xi,\zeta)+\big[\Theta(\xi),\Theta(\zeta)\big]=0
\label{s91}
\ee
where $\text{d}$ is the exterior derivative on $G$ and $[\cdot,\cdot]$ is the Lie bracket of $\mg$. As a result, one might think that the curvature of (\ref{ss84}) similarly vanishes, which would imply that the holonomy (\ref{ss87}) is trivial. However, this naive expectation is misguided: while the gauge connection (\ref{ss84}) does contain the Maurer-Cartan form, it also crucially contains the sandwiching within $\bra\phi_i|...|\phi_j\ket$ whose effect is to project the Maurer-Cartan form on the Lie algebra of the stabilizer; this projected form does {\it not}, in general, have a vanishing curvature. We shall confirm this explicitly in section \ref{sec6} with the example of the Poincar\'e group.\\

A final comment concerns the space of parameters leading to the Berry holonomies (\ref{ss87}), which we originally introduced by considering closed paths in $G$. However, this point of view is somewhat too restrictive: since the states $|\phi_1\ket$,...,$|\phi_1\ket$ have a non-trivial stabilizer (\ref{s88}), one is free to consider a path $f(t)$ in $G$ which only closes up to some element of $G_{\vec\phi}$, \ie $f(T)=f(0)h$ with $h\in G_{\vec\phi}$. Its projection on the quotient space $G/G_{\vec\phi}$ is closed and this is enough to ensure that the corresponding Berry phases are well-defined. Concretely, let us assume that the stabilizer is connected; then, for any open path $f(t)$ such that $f(T)=f(0)h$ with $h\in G_{\vec\phi}$, we define a {\it closed} path
\be
\bar f(t)
=
\begin{cases}
f(t) & \text{for }0\leq t\leq T\\
f(T)h(t) & \text{for }T\leq t\leq T'
\end{cases}
\label{s93}
\ee
where $h(t)$ is a curve in $G_{\vec\phi}$ such that $h(T)=e$ and $h(T')=f(T)^{-1}f(0)$. Formula (\ref{ss87}) applies to the path $\bar f$, whose Berry holonomy factorizes as
\be
B_{\vec\phi}[\bar f]
=
P\exp\Big[-\int_h\bra\vec\phi|\cu[\Theta]|\vec\phi\ket^t\Big]\cdot P\exp\Big[-\int_f\bra\vec\phi|\cu[\Theta]|\vec\phi\ket^t\Big].
\nn
\ee
Here the first term, due to the stabilizer path $h(t)$, only depends on its endpoints and does not depend on the choice of $h(t)$. Accordingly, we {\it define} the Berry holonomy of a (generally open) path $f(t)$ such that $f(T)^{-1}f(0)\in G_{\vec\phi}$ as follows:
\be
B_{\vec\phi}[f]
\equiv
\bra\vec\phi|\cU[f(T)^{-1}f(0)]|\vec\phi\ket^tP\exp\Big[-\int_f\bra\vec\phi|\cu[\Theta]|\vec\phi\ket^t\Big].
\label{ss94}
\ee
This says that the actual space of parameters is the coset space $G/G_{\vec\phi}$. It implies that, given a path $f(t)$, it can have non-zero Berry phases only if its projection on $G/G_{\vec\phi}$ is closed and contains more than one point. This is consistent with the `orbit method' for building group representations \cite{woodhouse1997}, where geometric quantization of a coadjoint orbit $G/G_{\vec\phi}$ produces a unitary representation of $G$; from that perspective, Berry phases associated with loops in $G$ coincide with symplectic fluxes on $G/G_{\vec\phi}$ \cite{boya2001berry}. We will encounter similar observations below for semi-direct products.

%%%%%%%%%%%%%%%%%%%%%%%%%%%%%%%%%%%%%
\subsection{Wigner-Berry phases}
\label{sec52}
%%%%%%%%%%%%%%%%%%%%%%%%%%%%%%%%%%%%%

Having reviewed some aspects of Berry phases, we now return to the original setting of this paper and consider a semi-direct product $G\ltimes A$; we also let $\cU$ be a Wigner-Mackey representation (\ref{s19}) specified by a momentum orbit $\cO_p$ and a spin representation $\cS$. As before we can see $G\ltimes A$ as a group of transformations relating various reference frames: $G$ consists of rotations and boosts, while $A$ consists of translations. We assume that $A$ contains a one-parameter subgroup of time translations, as is indeed the case for Poincar\'e, Bargmann and BMS groups. Then, if the system is prepared in an eigenstate of the Hamiltonian and if the reference frame changes adiabatically and returns to its initial configuration after some time, the final state vector should contain Berry phase factors. Our goal is to understand how those phases can be evaluated by adapting eq.\ (\ref{ss94}) to semi-direct products; the result will crucially involve Wigner rotations. Along the way we will encounter several technical complications, some of which we will not address rigorously. To streamline the presentation, a more detailed discussion of some of these issues is postponed to section \ref{sec54}. Accordingly, one can think of the next few pages as an intuitive motivation for the construction of the connection one-form of eq.\ (\ref{ss25b}) below; this one-form can be studied in its own right, irrespective of its representation-theoretic origin, and we will indeed see in section \ref{sec53} that it has many interesting properties.

\paragraph{Energy eigenstates.} At the outset, one should understand what is meant by `time translations': we are assuming that, in a certain reference frame, a vector $\alpha_0\in A$ generates time translations in the sense that a time translation by $t\in\RR$ is a group element $(e,t\alpha_0)\in G\ltimes A$, where $e$ is the identity in $G$. In another frame, related to the original one by a `boost' $f$ say, the same translation would be seen as $(f,0)\cdot(e,t\alpha_0)\cdot(f,0)^{-1}=(e,t\sigma_f\alpha_0)$. This is a semi-direct product analogue of the statement surrounding eq.\ (\ref{s81}).\\

In this language, an energy eigenstate (in the original frame) is a wavefunction $\Psi$ in the Hilbert space of $\cU$ that transforms under time translations as $\cU[(e,t\alpha_0)]\Psi=e^{-iEt}\,\Psi$. Given such a wavefunction, the boosted state $\cU[(f,0)]\Psi$ has energy $E$ with respect to time translations generated by $\sigma_f\alpha_0$. Since $\cU$ is given by the Wigner-Mackey formula (\ref{s19}), this means that $\Psi$ must have at least one definite component of momentum --- its energy. In particular, one may consider wavefunctions with definite momentum $k\in\cO_p$,
\be
\Psi(q)
=
\delta_k(q)v
\equiv
\Psi_{k,v}(q),
\label{ss20}
\ee
where $v\in\mh$ is a spin vector while $\delta_k$ is the Dirac distribution at $k\in\cO_p$ associated with the measure $\mu$ of (\ref{ss19}). Such plane waves are energy eigenstates by construction, for any $k$ and any $v$. Accordingly, from now on we assume that energy (with respect to $\alpha_0$) is bounded from below on the orbit $\cO_p$ and that its minimum is reached at $p$, as is indeed the case for the massive particles described in section \ref{sec3}. One can then think of $p$ as the energy-momentum vector of a particle in the rest frame, so plane waves $\Psi_{p,v}$ given by (\ref{ss20}) describe the possible states of a particle at rest. There are in general many such rest-frame states, since the particle may have a non-trivial spin space $\mh$.\\

At this point we must face a first technical subtlety: if the energy function on the orbit is non-constant, wavefunctions with definite energy must contain a delta function in momentum space, which implies that they are not square-integrable and do not belong to the Hilbert space. When it comes to Berry phases, this means that the treatment of section \ref{sec51} does not apply; instead we must consider superpositions of energy eigenstates, whose Berry phases are not sharply defined but satisfy a certain probability distribution \cite{Wu}. But for the sake of simplicity (and at the expense of rigour), we will adopt a somewhat heuristic viewpoint and consider normalized linear combinations of plane waves whose spread can be made arbitrarily small, in such a way that they formally approach energy eigenstates in the limit of zero spread. One can typically choose such smeared wavefunctions to be Gaussian coherent states. This point of view allows us to think of any normalized plane wave $\Phi_{k,v}$ with momentum $k$ (and unit spin vector $v\in\mh$) as a limit
\be
\Phi_{k,v}=\lim_{\lambda\rightarrow 0}\Phi_{k,v}^{\lambda}
\label{s99}
\ee
where $\Phi_{k,v}^{\lambda}$ is a normalized wavefunction peaked at $k$ with spread $\lambda$ in momentum space. Note that the resulting plane wave $\Phi_{k,v}$ is {\it not} of the form (\ref{ss20}) because the Dirac delta function is not normalized. Instead, the relation between the normalized plane wave (\ref{s99}) and the non-normalized one (\ref{ss20}) is formally $\Phi_{k,v}=\Psi_{k,v}/\sqrt{\delta_k(k)}$, where $\delta_k(k)$ is an infrared-divergent delta function evaluated at zero momentum. In Poincar\'e one would find $\delta_k(k)\propto\delta^{(D)}(0)$ where $\delta^{(D)}$ is the standard Dirac distribution on $\RR^D$; in a large volume $V$, $\delta_k(k)\propto V$. With this prescription for regulating infrared divergences, one has scalar products such as $\bra\Phi_{k,v}|\Phi_{k,w}\ket=(v|w)$. In particular, writing $\text{dim}(\mh)=N$ and letting $v_1,v_2,...,v_N$ be an orthonormal basis of $\mh$, we get $N$ linearly independent vectors
\be
\Phi_{p,i}
=
v_i\frac{\delta_p}{\sqrt{\delta_p(p)}}
\qquad
\text{such that}
\qquad
\bra\Phi_{p,i}|\Phi_{p,j}\ket=\delta_{ij}.
\label{ss101}
\ee
One can think of the $\Phi_{p,i}$'s as analogues of highest-weight states. Boosted vectors such as $\cU[(f,\alpha)]\Phi_{p,i}$ can then be seen as linear combinations of `descendant states'; even though the energy spectrum is generally continuous (think \eg of the relativistic energy $\sqrt{M^2+\bk^2}$), the adiabatic theorem applies \cite{Avron:1998th} and the considerations of section \ref{sec51} suggest that loops in $G\ltimes A$ lead to Berry phases.

\paragraph{Berry phases and Maurer-Cartan form.} Consider a closed path $(f(t),\alpha(t))$ of reference frame transformations in $G\ltimes A$. Let $\cU[(f(0),\alpha(0))]\Phi_{p,i}$ be the initial state vector of the system and let it evolve according to the time-dependent Schr\"odinger equation (\ref{ss83}):
\be
i\der_t\Psi
=
\cU\big[\big(f(t),\alpha(t)\big)\big]\,i\cu[(0,\alpha_0)]\,\cU\big[\big(f(t),\alpha(t)\big)\big]^{-1}\cdot\Psi.
\nn
\ee
If the path $(f(t),\alpha(t))$ is traced very slowly, the adiabatic theorem \cite{Avron:1998th} ensures that the wavefunction $\Psi(t)$ at time $t$ is given by an expression similar to eq.\ (\ref{s83}). Once the path $(f(t),\alpha(t))$ closes, say at $t=T$, the wavefunction $\Psi(T)$ differs from $\cU[(f(0),\alpha(0))]\Phi_{p,i}$ by a Berry holonomy (\ref{ss87}) that now takes the form
\be
B_{\vec\Phi_p}\big[\big(f(t),\alpha(t)\big)\big]
=
P\exp\bigg[-\oint_{(f,\alpha)}\big<\vec\Phi_p\big|\cu[\vartheta]\vec\Phi_p^t\big>\bigg]
\label{s103}
\ee
where $\vec\Phi_p$ denotes the column vector of wavefunctions $\big(\Phi_{p,1},\Phi_{p,2},...,\Phi_{p,N}\big)^t$ while $\cu$ is the Lie algebra representation (\ref{ss47}) and $\vartheta$ is the Maurer-Cartan form of $G\ltimes A$. To compute the latter, consider a path $(g(t),\beta(t))$ in $G\ltimes A$ such that $g(0)=f$, $\beta(0)=\alpha$. This defines a tangent vector $(\dot g(0),\dot\beta(0))\in T_{(f,\alpha)}\big(G\ltimes A\big)\cong T_fG\oplus A$. The Maurer-Cartan form at $(f,\alpha)$ acting on that tangent vector is
\be
\left.\frac{d}{dt}\right|_{0}\Big[(f,\alpha)^{-1}\cdot\big(g(t),\beta(t)\big)\Big]
\refeq{s13}
\Big(\left.\frac{d}{dt}\right|_{0}\big(f^{-1}\cdot g(t)\big),\sigma_{f^{-1}}\dot\beta(0)\Big).
\nn
\ee
Here the first entry is the Maurer-Cartan form of $G$ acting on $\dot g(0)\in T_fG$. Thus the Maurer-Cartan form of $G\ltimes A$ is $\vartheta_{(f,\alpha)}=\big(\Theta_f,\sigma_f^{-1}\big)$, where the two entries respectively act on $T_fG$ and $T_{\alpha}A=A$.\footnote{The same Maurer-Cartan form recently appeared in \cite{Barnich:2017jgw} in the context of BMS$_3$ symmetry.} It follows that (\ref{s103}) can be written as
\be
B_{\vec\Phi_p}\big[\big(f(t),\alpha(t)\big)\big]
=
P\exp\bigg[-\oint_{(f,\alpha)}\big<\vec\Phi_p\big|\cu[(\Theta,\sigma^{-1})]\vec\Phi_p^t\big>\bigg]
\label{ss105}
\ee
and it remains to put this in a simpler form by massaging the Berry connection in the argument of the exponential. To do this we treat separately the `rotational piece' $\Theta$ and the `translational piece' $\sigma^{-1}$.

\paragraph{Translational piece.} Consider the contribution of the translational Berry phase, due to the term involving $\alpha$ in the representation (\ref{ss47}). From eq.\ (\ref{ss105}) and the scalar product (\ref{ss101}) we find that this term contributes an overall (Abelian) phase
\be
B^{\text{transl.}}_{\vec\Phi_p}\big[(f,\alpha)\big]
=
P\exp\Big[-\oint_{(f,\alpha)}\big<\vec\Phi_p\big|\cu[(0,\sigma^{-1})]\vec\Phi_p^t\big>\Big]
\refeq{ss47}
\exp\Big[-i\oint_0^T\!\!\text{d}t\,\langle p,\sigma_{f(t)}^{-1}\dot\alpha(t)\rangle\Big]
\label{BerryT}
\ee
where we neglect to write an identity operator $\II\in\text{End}(\mh)$ on the right-hand side. Using the definition (\ref{s15}) of the action of $G$ on momenta, this can be rewritten as
\be
B^{\text{transl.}}_{\vec\Phi_p}\big[(f,\alpha)\big]
=
\exp\Big[-i\oint_0^T\!\!\text{d}t\,\langle f(t)\cdot p,\dot\alpha(t)\rangle\Big].
\label{trabi}
\ee
This overall phase vanishes whenever the path $\alpha(t)$ is constant or when $f(t)$ is contained in the little group. But in general it produces a non-zero contribution to the total Berry holonomy (\ref{ss105}), albeit one that is insensitive to spin; in particular, it also affects scalar particles. One can think of it as a symplectic flux on the cotangent bundle $T^*\cO_p$, when the latter is endowed with its usual symplectic form; indeed $f(t)\cdot p=q(t)$ is a closed path on $\cO_p$ (`momentum space') while $\alpha(t)$ is a closed curve in position space, and the exponent of (\ref{trabi}) can be seen as the integral of the Liouville one-form (the symplectic potential) on $T^*\cO_p$ along the path $\big(q(t),\alpha(t)\big)$. This is consistent with the results of \cite{boya2001berry} and also with the general relation between holonomies on homogeneous spaces and symplectic fluxes (see \eg \cite{neeb2002central}), although we are not aware of any reference that describes analogous observations for semi-direct products. In the context of Thomas precession, the phase (\ref{trabi}) is generally neglected, precisely because it is blind to spin; for the same reason, from now on we let the translational path $\alpha(t)$ be constant ($\dot\alpha=0$) so that the only non-zero contribution to the Berry holonomy (\ref{ss105}) comes from its rotational piece.

\paragraph{Rotational piece.} When $\alpha(t)$ is constant, the Berry phases of (\ref{ss105}) are entirely due to the path $f(t)\in G$, whose contribution to the integrand in the exponent of (\ref{ss105}) is
\be
\big<\vec\Phi_p\big|\cu[(\Theta_f,0)]\vec\Phi_p^t\big>
\refeq{ss47}
\Big<\vec\Phi_p\Big|
\Big(\cs\big[\Ad_{g_p^{-1}}\Theta_f-(g^*\Theta)_p(\xi_{\Theta_f})_p\big]-\big(\xi_{\Theta_f}\big)_p\Big)
\vec\Phi_p^t\Big>.
\label{s107}
\ee
To simplify this expression we use eq.\ (\ref{ss101}) for the states $\Phi_{p,i}$. In particular, the vector field outside the argument of $\cs$ in (\ref{s107}) is blind to the spin vector $v_i$ of $\Phi_{p,i}$ and contributes a term $\bra\vec\Phi_p|(\xi_{\Theta_f})_p\vec\Phi_p^t\ket$, which is proportional to $\bra\delta_p|(\xi_{\Theta_f})_p\delta_p\ket$. This is the expectation value of a boost generator in a state at rest, and therefore vanishes; intuitively, the boosted state $\xi_X\Phi_p$ is orthogonal to $\Phi_p$ because it has a different value of angular momentum \cite{Campoleoni:2016vsh}. Thus the only non-zero piece of (\ref{s107}) comes from the Wigner generator $\cs[...]$; the latter is only sensitive to the spin of $\Phi_{p,i}$, and the scalar product (\ref{ss101}) yields
\be
\Big<\vec\Phi_p\Big|
\cs\Big[\Ad_{g_p}^{-1}\Theta_f-(g^*\Theta)_p\big(\xi_{\Theta_f}\big)_p\Big]\vec\Phi_p^t\Big>
=
\cs\Big[\Ad_{g_p}^{-1}\Theta_f-(g^*\Theta)_p\big(\xi_{\Theta_f}\big)_p\Big]
\label{s25b}
\ee
where we identify the operator $\cs[...]$ with its matrix in the basis $v_1,...,v_N$ of $\mh$. This is the simplification we were looking for; we now analyse it in detail.

%%%%%%%%%%%%%%%%%%%%%%%%%%%%%%%%%%%%%
\subsection{Thomas precession}
\label{sec53}
%%%%%%%%%%%%%%%%%%%%%%%%%%%%%%%%%%%%%

Formula (\ref{s25b}) is the main result of this paper. It is a Wigner generator (\ref{s47}) that coincides with the Berry connection associated to adiabatic boosts and rotations,
\be
\boxed{A
=
\cs\Big[\Ad_{g_p}^{-1}\Theta-(g^*\Theta)_p\big(\xi_{\Theta}\big)_p\Big]
\refeq{s47}
\sfw_p[\Theta],}
\label{ss25b}
\ee
and it may be seen as (the representation $\cs$ of) the projection of the Maurer-Cartan form $\Theta$ on the little algebra. We shall refer to it as a {\it Wigner-Berry connection}. It can be simplified by choosing a gauge for standard boosts; specifically, let us assume that $g_p=e$ is the identity as in section \ref{sec43}. Then $\Ad_{g_p^{-1}}\Theta=\Theta$ and the pullback of the Maurer-Cartan form by $g$ boils down to $(g^*\Theta)_p=\Theta_{g_p}\text{d}g_p=\text{d}g_p$. As a result (\ref{ss25b}) becomes
\be
\boxed{\Big.A
=
\cs\big[\Theta-\text{d}g_p(\xi_{\Theta})_p\big]}
\qquad
\text{(for $g_p=e$).}
\label{s109}
\ee
We will see below that this is essentially the connection one-form anticipated in (\ref{s67}).\\

The observations of the last few pages show that Wigner rotations, Thomas precession and Berry phases are one and the same thing when it comes to Wigner-Mackey representations. Indeed, the Berry holonomy (\ref{ss105}) of (\ref{s109}) along a loop $f(t)$ is
\be
B_{\vec\Phi_p}[f(t)]
=
P\exp\Big[-\oint_f\cs\big[\Theta-\text{d}g_p(\xi_{\Theta})_p\big]\Big]
\qquad\text{(for $g_p=e$).}
\label{s109b}
\ee
This is typically a rotation matrix that may be interpreted as the net precession of a particle's spin obtained by superimposing a sequence of infinitesimal Wigner rotations. Similar statements have already appeared in the literature \cite{Mathur:1991ynd,Aravind,Brezov:2015aja,Stone:2014fja}, but to the best of our knowledge, all of them only deal with the Poincar\'e group. By contrast, our goal here was to draw general conclusions for arbitrary semi-direct products. In particular, we now know that Thomas precession occurs if and only if the curvature of the connection (\ref{ss25b}) does not vanish identically. As argued around eq.\ (\ref{s91}), this curvature need not vanish thanks to the fact that it is a {\it projection} of the Maurer-Cartan form down to the stabilizer.\\

A comment is in order about the parameter space on which the connection (\ref{s109}) lives. Indeed, note that for any path $f(t)$ entirely contained in $G_p$, the Berry holonomy (\ref{s109b}) is trivial; this is because in that case the vector field $\xi_{\Theta_f(\dot f)}$ vanishes at $p$ and the remaining connection $\sfw_p[\Theta]\big|_{G_p}=\cs[\theta]$ is flat due to the Maurer-Cartan equation (\ref{s91}). One can thus think of the parameter space of the system as being not the group manifold $G$, but the coset space $G/G_p\cong\cO_p$, and the only thing that truly matters is the {\it projection} of $f(t)$ on $\cO_p$, \ie the path $\gamma(t)=f(t)\cdot p$ in momentum space. When that projection is a closed curve that contains more than one point, the Berry holonomy (\ref{s109b}) is generally non-zero. (We encountered a similar observation around eq.\ (\ref{s93}).) Furthermore, for any such path $f(t)$, whenever $f(T)=f(0)\cdot h$ for some group element $h\in G_p$, one can define a holonomy (\ref{ss94}) even though the path $f$ is not closed.\\

This implies that curves which only consist of standard boosts contain all the information about the Berry phases associated with paths in $G$. In other words, as far as Thomas precession is concerned we are free to consider without loss of generality only paths of the form $f(t)=g_{q(t)}$, where $q(t)$ is some closed curve in $\cO_p$. Any such path can be interpreted as the momentum-space trajectory $q(t)$ of a particle. Its `acceleration' (or rather the force acting on the particle) is the tangent vector $\dot q(t)$, and the pullback $\cA=g^*A$ of the Berry connection (\ref{s109}) acting on that vector is
\begin{align}
\cA_{q(t)}(\dot q(t))
&=
\cs\Big[\Theta_{g_{q(t)}}\text{d}g_{q(t)}\dot q(t)-\text{d}g_p\big(\xi_{\Theta_{g_q(t)}\text{d}g_{q(t)}\dot q(t)}\big)_p\Big]\nn\\
\label{s113}
&= \cs\big[g^*\Theta-\text{d}g_p\text{d}(g^{-1}\cdot)\big]_{q(t)}(\dot q(t)).
\end{align}
Here we recognize the one-form (\ref{s67}) of section \ref{sec43}, so the lemma surrounding (\ref{s69}) applies and we conclude that, if all Wigner rotations (\ref{s20}) associated with standard boosts are trivial (\ie if condition (\ref{s69}) holds), then the Berry connection $\cA$ in (\ref{s113}) vanishes identically. In particular, the triviality of Wigner rotations associated with standard boosts implies the absence of Berry phases and Thomas precession. Conversely, if some Berry holonomies (\ref{s109}) are non-trivial, then some Wigner rotations of standard boosts must also be non-trivial.

%%%%%%%%%%%%%%%%%%%%%%%%%%%%%%%%%%%%%
\subsection{Technical remarks}
\label{sec54}
%%%%%%%%%%%%%%%%%%%%%%%%%%%%%%%%%%%%%

Here we discuss some of the technical issues encountered in section \ref{sec52}. A first remark concerns the Berry phase (\ref{BerryT}) associated with non-constant translations. Namely, if the complete parameter space of the system truly was $G/G_p=\cO_p$, as is the case for pure Thomas precession, then translations would not contribute to Wigner-Berry phases. But one should keep in mind that $\cO_p$ is only the parameter space that incorporates the effects of pure boosts and rotations; by contrast, the full parameter space of the system is actually a coadjoint orbit of $G\ltimes A$, and is always larger than $\cO_p$. For scalar particles this orbit is a cotangent bundle $T^*\cO_p$ in which $\cO_p$ is the parameter space for rotations and boosts, while the cotangent piece forms a parameter space for translations. For spinning particles the situation is similar, but the coadjoint orbit has a more intricate structure: it is a bundle of little group coadjoint orbits over $T^*\cO_p$ --- see \cite{Rawnsley1975} for the original derivation and \cite{Barnich:2015uva} or \cite[sec.\ 5.4]{Oblak:2016eij} for more recent presentations. This is why the symplectic form of $T^*\cO_p$ appears in (\ref{trabi}), and it explains why translations can lead to non-vanishing Berry phases despite being blind to spin. It is also consistent with our remark below (\ref{ss94}) on the relation between Berry phases in group representations and coadjoint orbits.\\

Note that this conclusion, though correct, is at odds with our arguments at the end of section \ref{sec51} based on the stabilizer (\ref{s88}). Indeed, the stabilizer of the rest-frame states (\ref{ss101}) in the sense of (\ref{s88}) is the entire group $G_p\ltimes A$, which naively predicts that the parameter space is $(G\ltimes A)/(G_p\ltimes A)\cong\cO_p$. By contrast, the stabilizer of a coadjoint orbit of $G\ltimes A$ is $G_p\ltimes A_p$, where $A_p$ is a strict subspace of $A$. (For example, for massive particles $A_p\cong\RR$ typically consists of pure time translations.) This distinction between $A$ and $A_p$ is crucial, as it accounts for the cotangent piece of scalar coadjoint orbits $(G\ltimes A)/(G_p\ltimes A_p)\cong T^*\cO_p$, and thus leaves room for translational Berry phases (\ref{trabi}). So the stabilizer (\ref{s88}) misses a key restriction on translations when it is bluntly applied to Wigner-Mackey representations. The reason for this failure is that the rest-frame states (\ref{ss101}) do not, strictly speaking, belong to the Hilbert space, hence cannot be used as harmlessly as the states $|\phi_i\ket$ of section \ref{sec51}.\\

Finally, note that in writing (\ref{s103}) we relied on the adiabatic theorem for systems without gap in the energy spectrum \cite{Avron:1998th}. This is indeed needed for Wigner-Mackey representations since they generally have continuous energy, but in practice one may not have to address this subtlety at all. Indeed, any particle that undergoes periodic changes of reference frames (\eg an electron bound to an atomic nucleus) is actually located in an attractive potential, so that its possible energy levels {\it are} discrete after all. Of course, this discreteness becomes visible only with a dynamical analysis that goes beyond the kinematical treatment of this paper.\\

To conclude we should mention one alternative point of view that might justify our derivation of (\ref{ss25b}) without invoking coherent or bound states. As mentioned earlier, most Wigner-Mackey representations of semi-direct products can be obtained by quantizing their coadjoint orbits \cite{Rawnsley1975,Li1993}. In that context the classification of these representations in terms of orbits and spins can be seen as a quantum version of the classification of coadjoint orbits. Each orbit is the phase space of a (generally spinning) particle. Furthermore, since by assumption the symmetry group contains time translations, the system comes equipped with a whole family of Hamiltonian functions that are related to one another by changes of reference frames. (This statement is the classical counterpart of eq.\ (\ref{s81}).) One can thus think of the set of inequivalent reference frames of the particle as a space of parameters, and look for the Hannay angles \cite{Hannay} that appear when these parameters are slowly varied in a cyclic way. Since they are purely classical objects, these angles are not plagued by the difficulties listed above: as there are no `wavefunctions', there are no issues with the normalizability of quantum states or the adiabatic theorem. Our expectation is that the Berry connection (\ref{ss25b}) is a quantum analogue of a classical Hannay connection on phase space, but we will not attempt to verify this here.

%%%%%%%%%%%%%%%%%%%%%%%%%%%%%%%%%%%%%%%%%%%%%%%%%%%%%%%%%%%
\section{Applications and outlook}
\label{sec6}
%%%%%%%%%%%%%%%%%%%%%%%%%%%%%%%%%%%%%%%%%%%%%%%%%%%%%%%%%%%

In this short and last section, we illustrate the results of section \ref{sec5} with the Poincar\'e and Bargmann groups and briefly discuss other cases where similar tools apply.

%%%%%%%%%%%%%%%%%%%%%%%%%%%%%%%%%%%%%
\subsection{Thomas precession for relativistic particles}
\label{sec61}
%%%%%%%%%%%%%%%%%%%%%%%%%%%%%%%%%%%%%

Relativistic and non-relativistic particles are irreducible unitary representations of the Poincar\'e and Bargmann groups, respectively. As in section \ref{sec3} we only consider massive representations in $D+1$ dimensions. Let us start with the non-relativistic case --- the Bargmann group. As we showed in (\ref{s41}), the Wigner rotations associated with standard boosts are trivial, so the discussion below (\ref{s113}) implies that there is no Thomas precession.  This can be confirmed by directly evaluating the Berry connection (\ref{s109}), or rather its pullback (\ref{s113}), for the Bargmann group: since standard boosts are given by (\ref{s37}), they pullback the Maurer-Cartan form according to $(g^*\Theta)_q=g_q^{-1}\text{d}g_q=(0,\text{d}\bbq/M)$, where $\text{d}\bbq$ is an $\RR^D$-valued one-form on $\RR^D$, while the entire one-form $g^*\Theta$ takes its values in the Euclidean Lie algebra $\mathfrak{so}(D)\inplus\RR^D$. The second one-form appearing in (\ref{s113}) is also $\text{d}g_p\text{d}(g^{-1}\cdot)_q=(0,\text{d}\bbq/M)$, so the Berry gauge field of (\ref{s113}) vanishes.\\

Relativistic particles are much more interesting in that respect. Consider a particle with spin $\cS$, the latter being some irreducible unitary representation of the little group $\text{SO}(D)$. Standard boosts are given by (\ref{s31}), and their Wigner rotations are (\ref{s33}). They are non-trivial, so one may expect relativistic spinning particles to be subject to Thomas precession. To confirm this, consider the argument of the Berry connection (\ref{s113}),
\be
\cA_q
\equiv
g_q^{-1}\text{d}g_q-\text{d}g_p\text{d}(g_q^{-1}\cdot)_q.
\label{s126}
\ee
The first term of this expression is just the pullback of the Maurer-Cartan form of the Lorentz group by the standard boosts (\ref{s31}), while the second projects the Maurer-Cartan form down to the little algebra. Indeed, using (\ref{s31}) we explicitly find
\be
g_q^{-1}\text{d}g_q
=
\begin{pmatrix}
0 & \frac{\text{d}\bbq^t}{M}-\Big(1-\frac{1}{\sqrt{1+\bbq^2/M^2}}\Big)\frac{\bbq^t\text{d}\bbq}{\bbq^2}\frac{\bbq^t}{M}\\[.5cm]
\frac{\text{d}\bbq}{M}-\Big(1-\frac{1}{\sqrt{1+\bbq^2/M^2}}\Big)\frac{\bbq^t\text{d}\bbq}{\bbq^2}\frac{\bbq}{M} & \big(\sqrt{1-\bbq^2/M^2}-1\big)\Big(\frac{\text{d}\bbq\bbq^t-\bbq \text{d}\bbq^t}{\bbq^2}\Big)
\end{pmatrix},
\label{ss126}
\ee
which is valued in the Lorentz algebra as it should. Its purely spatial components $(i,j)$ span a matrix in $\mathfrak{so}(D)$, while its first row and first column generate boosts. Crucially, that boost piece does not appear in the connection (\ref{s126}) because it is cancelled by
\be
\text{d}g_p\text{d}(g_q^{-1}\cdot)_q
=
\begin{pmatrix}
0 & \frac{\text{d}\bbq^t}{M}-\Big(1-\frac{1}{\sqrt{1+\bbq^2/M^2}}\Big)\frac{\bbq^t\text{d}\bbq}{\bbq^2}\frac{\bbq^t}{M}\\[.3cm]
\frac{\text{d}\bbq}{M}-\Big(1-\frac{1}{\sqrt{1+\bbq^2/M^2}}\Big)\frac{\bbq^t\text{d}\bbq}{\bbq^2}\frac{\bbq}{M} & 0
\end{pmatrix}.
\nn
\ee
As a result one finds that the connection (\ref{s126}) is just the rotational piece of (\ref{ss126}),
\be
\cA_q
=
\begin{pmatrix}
0 & 0\\
0 & \ds\big(\sqrt{1+\bbq^2/M^2}-1\big)\Big(\frac{\text{d}\bbq\bbq^t-\bbq \text{d}\bbq^t}{\bbq^2}\Big)
\end{pmatrix}.
\label{s127}
\ee
It is non-zero, consistently with the fact that the Wigner rotations (\ref{s33}) are non-trivial. To prove that Thomas precession exists, it remains to verify that the connection (\ref{s127}) is not flat. We do this by choosing a basis of the little algebra $\mathfrak{so}(D)$: for $1\leq i<j\leq D$ we define a $D\times D$ matrix $T_{ij}$ having an entry $1$ in the $j^{\text{th}}$ row and $i^{\text{th}}$ column, $(-1)$ in the $i^{\text{th}}$ row and $j^{\text{th}}$ column, and zero elsewhere; in components, $(T_{ij})_{kl}=\delta_{il}\delta_{jk}-\delta_{ik}\delta_{jl}$. There are $D(D-1)/2$ such matrices generating the $\mathfrak{so}(D)$ algebra, and their commutators read
\be
[T_{ij},T_{kl}]
=
\delta_{ik}T_{jl}-\delta_{il}T_{jk}-\delta_{jk}T_{il}+\delta_{jl}T_{ik},
\label{s7r}
\ee
where $T_{ji}\equiv-T_{ij}$ if $i<j$. We can then write the connection (\ref{s127}) and its curvature $\cF$ as $\cA=\cA_{ij}T_{ij}$ and $\cF=\cF_{ij}T_{ij}$, with implicit summation over $i<j$. Using (\ref{s7r}) one finds
\be
(\cF_{ij})_q
=
\text{d}(\cA_{ij})_q+[\cA_q,\cA_q]_{ij}
=
\demi\frac{\text{d}q_i\wedge \text{d}q_j}{M^2}
-
\frac{1}{2M^4}\frac{q_k\text{d}q_k\wedge q_{[i}\text{d}q_{j]}}{1+\bbq^2/M^2+\sqrt{1+\bbq^2/M^2}}
\label{ss127}
\ee
where $q_{[i}\text{d}q_{j]}\equiv q_i\text{d}q_j-q_j\text{d}q_i$. This does not vanish, so the Berry curvature associated with changes of reference frames in (massive) representations of the Poincar\'e group is non-zero: relativistic particles with non-zero spin are subject to Thomas precession.\\

It is worth comparing (\ref{s127}) and (\ref{ss127}) with the expressions for Thomas precession that can be found in the literature (see \eg \cite[sec.\ 11.8]{Jackson:1998nia} or \cite{Dragan:2012ir}). For instance, the Berry connection (\ref{s127}) is essentially a higher-dimensional cross product,
\be
\big(\sqrt{1+\bbq^2/M^2}-1\big)\frac{\text{d}\bbq\times\bbq}{\bbq^2},
\label{s128}
\ee
and may be seen as the infinitesimal rotation that results from a boost with rapidity $\text{d}\bbq/M$ applied to a particle with momentum $\bbq$. In terms of the velocity $\bv$ such that $\bbq=M\bv/\sqrt{1-\bv^2/c^2}$, the rotation generator (\ref{s128}) is
\be
\bigg(\frac{1}{\sqrt{1-\bv^2/c^2}}-1\bigg)\frac{\text{d}\bv\times\bv}{\bv^2}
\,\:\;\stackrel{\bv/c\rightarrow0}{\sim}\;\;\,
\demi\,\frac{\text{d}\bv\times\bv}{c^2}
\nn
\ee
where we reinstate the speed of light $c$ for convenience. The right-hand side here is the first term of the non-relativistic expansion of the exact result, and contains the notorious `Thomas half' \cite{Thomas:1926dy} that reconciled the observations of Uhlenbeck and Goudsmit \cite{Uhlenbeck} with the predictions of quantum mechanics. In that context the differential form $\text{d}\bv$ is interpreted as a small variation of a particle's velocity due to its acceleration --- for instance the centripetal acceleration of an electron bound to an atomic nucleus.\\

Of course, any statement based solely on the connection (\ref{s127}) is bound to be gauge-dependent, hence unobservable; for instance, we would have obtained a different expression for the rotation generator (\ref{s128}) if we had used different standard boosts. The only truly observable quantities are gauge-invariant; those are typically the Berry phases that can be extracted from the holonomies of the connection (\ref{s127}) and that may be interpreted as rotations corresponding to the Thomas precession of a spinning particle. Note that in general (for $D\geq 3$) even the curvature (\ref{ss127}) is not directly observable; the only exception occurs in three space-time dimensions ($D=2$), where the little group $\text{SO}(2)$ is Abelian so that the curvature (\ref{ss127}) is gauge-invariant and reduces to
\be
\cF
=
\frac{1}{2M^2}\frac{\text{d}q_1\wedge \text{d}q_2}{\sqrt{1+\bbq^2/M^2}}
\begin{pmatrix} 0 & -1 \\ 1 & 0 \end{pmatrix}.
\nn
\ee
The two-form in front of the matrix may be interpreted as the volume form of a hyperbolic plane embedded in energy-momentum space, so that the net angle of rotation undergone by a particle that follows a closed path in rapidity space is the hyperbolic area of the enclosed surface (times the particle's spin) \cite{Aravind}.\\

The Berry connection (\ref{s127}) and its curvature (\ref{ss127}) have a number of interesting properties and may for instance be seen as solitonic configurations of a non-Abelian gauge field \cite{Mathur:1991ynd,Brezov:2015aja}. We will refrain from pursuing this line of thought here. Our main purpose in this work was indeed to point out that such a rich gauge-theoretic structure arises in the unitary representations of {\it any} semi-direct product.

%%%%%%%%%%%%%%%%%%%%%%%%%%%%%%%%%%%%%
\subsection{Dressed particles and other stories}
\label{sec62}
%%%%%%%%%%%%%%%%%%%%%%%%%%%%%%%%%%%%%

In this paper we have shown that a notion of `Thomas precession' exists for essentially all semi-direct product groups. This applies of course to Poincar\'e, but our approach is independent of the details of the group structure and thus allows us to be more general. A notable example is the BMS group, whose unitary representations are expected to describe particles dressed with soft gravitons. In that context it is natural to wonder if gravitational dressing contributes to Thomas precession, and if so, whether this contribution can actually be measured. We recently addressed a similar question in \cite{Oblak:2017ect} for the Virasoro group (see also \cite{Mao:2017axa} for related considerations in the gauge-theoretic realm); as for BMS representations, we intend to turn to them in a separate publication \cite{OblakThomas}.\\

While the BMS groups are undoubtedly interesting examples of groups whose representations display Thomas precession, they are by no means the only uncharted territory. For instance, the symmetry group of any two-dimensional conformal field theory with conserved currents is a semi-direct product between the Virasoro group and a Kac-Moody group, so the corresponding unitary representations may contain some sort of Thomas precession. (If the Kac-Moody algebra is non-Abelian the treatment of this paper does not apply, but this does not prevent the existence of Berry phases similar to those of section \ref{sec51}.) In that context an example that is both rich and tractable is the warped Virasoro group \cite{Detournay:2012pc}, which spans the symmetries of warped conformal field theories. We intend to investigate some of these questions in the future and hope that the present work can be useful for such considerations.

%%%%%%%%%%%%%%%%%%%%%%%%%%%%%%%%%%%%%%%%%%%%%%%%%%%%%%%%%%%
\section*{Acknowledgements}
\addcontentsline{toc}{section}{Acknowledgements}
%%%%%%%%%%%%%%%%%%%%%%%%%%%%%%%%%%%%%%%%%%%%%%%%%%%%%%%%%%%

I wish to thank G.\ Barnich, J.\ Fine and K.\ H.\ Neeb for stimulating discussions and correspondence. This work is supported by the Swiss National Science Foundation, and partly by the NCCR SwissMAP.

%\newpage
%\nocite{*}
\addcontentsline{toc}{section}{References}
%\bibliography{WignerBib}

\end{document}